\documentclass[11pt,a4paper]{article}

\usepackage[utf8]{inputenc}
\usepackage[T1]{fontenc}
\usepackage{lmodern}
\usepackage[a4paper,margin=1in]{geometry}
\usepackage{amsmath,amssymb,mathtools}
\usepackage{amsthm}
\usepackage{enumitem}
\usepackage{float}
\usepackage{microtype}
\usepackage[hidelinks]{hyperref}
\usepackage{color}
\usepackage{tikz}

\setlist[itemize]{leftmargin=1.4em,itemsep=0.25em,topsep=0.35em}
\setlist[enumerate]{leftmargin=1.6em,itemsep=0.25em,topsep=0.35em}

\newtheorem{theorem}{Theorem}[section]
\newtheorem{conjecture}[theorem]{Conjecture}

\theoremstyle{definition}

\theoremstyle{remark}

\newcommand{\dd}{\mathrm{d}}
\newcommand{\ZZ}{\mathbb Z}
\newcommand{\RR}{\mathbb R}
\newcommand{\NN}{\mathbb N}
\newcommand{\EE}{\mathbb E}
\newcommand{\PP}{\mathbb P}
\newcommand{\Var}{\operatorname{Var}}

\newcommand{\1}{\mathbf 1}

\newcommand{\Longto}{\Longrightarrow}
\newcommand{\PHN}{P_{\mathrm{HN}}}
\newcommand{\eps}{\varepsilon}
\newcommand{\Jdw}{\mathsf J_{\mathrm{dw}}}
\newcommand{\Jhom}{\mathsf J_{\mathrm{hom}}}
\newcommand{\SSEP}{\mathrm{SSEP}}

\hypersetup{
  pdftitle={Anomalous current fluctuations in the stochastic XNOR hopping model},
  pdfauthor={Balázs Pozsgay},
  pdfsubject={Stochastic XNOR, zero-range process, anomalous full counting statistics},
  pdfkeywords={XNOR, folded XXZ, full counting statistics, anomalous fluctuations, zero-range process, simple exclusion, half-normal distribution}
}

\begin{document}

\begin{center}
{\LARGE\bfseries Anomalous current fluctuations in the stochastic\\ XNOR hopping model}\\[0.35em]
{\large Bal\'azs Pozsgay}\\[0.25em]
{MTA-ELTE ``Momentum'' Integrable Quantum Dynamics Research Group}\\
{ELTE E\"otv\"os Lor\'and University, Budapest, Hungary}\\[0.35em]
{July 2026}\\[0.8em]
\end{center}

\begin{abstract}
We consider fluctuations of the spin current in the stochastic XNOR process, a kinetically constrained hopping
model in one spatial dimension.
We formulate three conjectures with explicit amplitudes for the long-time current distributions: a Gaussian limit on the
\(t^{1/4}\) scale for homogeneous initial states with nonzero magnetization, a half-normal limit on the \(t^{1/4}\) scale
for domain-wall states with opposite magnetizations, and an M-Wright limit on the \(t^{1/8}\) scale for homogeneous initial
states at zero magnetization.
Earlier work identified the tracer mechanism and anticipated the scaling exponents and limiting shapes in the domain-wall
and zero-magnetization settings. Starting from the microscopic XNOR dynamics, we predict the missing amplitudes for the
continuous-time  process and extend the picture to homogeneous biased initial data.
Simulations provide numerical support for all three conjectures without fitted parameters. A separate
mathematical companion paper presents an extensively AI-generated candidate proof. Generative-AI tools were used in the
research and writing workflow. 
\end{abstract}

\section{Introduction}
\label{sec:introduction}

This paper presents a statistical-physics analysis of current fluctuations in the stochastic XNOR process. It also
documents an extensive use of generative-AI systems in the analytic, computational, and writing workflow. We
first describe the scientific problem and the proposed results, and then state the role and verification status of the
AI-generated material. 

\subsection{The scientific problem}

Fluctuations in one-dimensional many-body systems can be much less conventional than their average transport laws suggest.  Even when a conserved density has the usual diffusive spreading length $t^{1/2}$, the time-integrated current through a fixed cut can exhibit unconventional statistics: its typical magnitude can have an anomalous time dependence, its limiting distribution can be non-Gaussian, and different cumulants can grow with different powers of time.  Such behaviour has been observed in integrable magnets and derived exactly in solvable deterministic and stochastic many-body dynamics \cite{KrajnikIlievskiProsen2022,KrajnikSchmidtPasquierIlievskiProsen2022,KrajnikSchmidtPasquierProsenIlievski2024,KrajnikKlobasBertiniProsen2025}.  Recent work has extended this picture to quantum many-body dynamics and has placed anomalous full counting statistics (FCS) in a broader hydrodynamic setting \cite{McCullochDeNardisGopalakrishnanVasseur2023,GopalakrishnanMcCullochVasseur2024,McCullochVasseurGopalakrishnan2025,Doyon2023,YoshimuraKrajnik2025,FujimotoEtAl2026}.
In particular, ballistic macroscopic fluctuation theory gives the anomalous equilibrium spin-current distribution of the easy-axis XXZ chain \cite{takato-stb-xxz-anomalous-fluct}.

A particularly transparent mechanism occurs in single-file systems with an inert colour.  The particles cannot pass one another, while the colour carried by each particle is transported by the motion but does not affect that motion.  The charge crossing a fixed cut is then a random sum over the particles swept past the cut by an effective tracer.  Two sources of fluctuations combine: the single-file displacement fixes the random number of terms, and the inert colours fix their signed sum.  This mechanism produces anomalous scaling and, when the mean colour vanishes, a genuinely non-Gaussian limit.  For several exactly solvable cellular automata these statements have been established directly from the microscopic dynamics, first in specific models and subsequently at the level of a charged single-file universality class \cite{KrajnikSchmidtPasquierIlievskiProsen2022,KrajnikSchmidtPasquierProsenIlievski2024,KrajnikKlobasBertiniProsen2025}.

The folded XXZ chain provides a natural quantum setting for the same physics.  Its hopping is kinetically constrained: a spin can move only when the surrounding spins have the appropriate configuration.  The constraint preserves a frozen background pattern and leads to Hilbert-space fragmentation, hard-rod kinematics, and an effective tracer description \cite{Zadnik2021,PozsgayEtAl2021,PozsgayGomborHutsalyuk2021,SalaEtAl2020,KhemaniHermeleNandkishore2020,MoudgalyaMotrunich2022}.  A stochastic version of this constrained hopping is the XNOR process studied here.  Its elementary move exchanges the middle spins of a four-site block only when the two outer spins agree.  An equivalent dynamics was studied earlier as a model of diffusing, reconstituting dimers \cite{MenonBarmaDhar1997}.

Subdiffusive XNOR transport was established in random-circuit dynamics by Singh, Ware, Vasseur, and Friedman, who found
the dynamical exponent $z=4$ \cite{SinghWareVasseurFriedman2021}.  Feldmeier, Witczak-Krempa, and Knap subsequently
derived an emergent single-file tracer description and the fully normalized long-time spin-correlation profile of the
random XNOR circuit, including its width \cite{FeldmeierWitczakKrempaKnap2022}.  These correlation-function results do
not directly determine the fixed-cut current law for homogeneous biased data in the continuous-time process considered here.

A complementary FCS analysis derived the domain-wall FCS for the folded XXZ setting, including the single-interface half-normal law at nonzero
magnetization and the equilibrium Gaussian variance-mixture law at zero magnetization
\cite{GopalakrishnanMorningstarVasseurKhemani2024,GopalakrishnanVasseur2023}. It also argued that the full generating function
carries over to stochastic XNOR after replacing the diffusive tracer scale by the subdiffusive $t^{1/4}$ scale, and gave
numerical support in the nonzero-magnetization case. Taken together, these results already anticipate the $t^{1/8}$ current scale and the same
mixture shape for stochastic XNOR at zero magnetization. They did not determine the microscopic current amplitudes for the
continuous-time XNOR process.

What was still missing was a fully normalized description of the current fluctuations for this precise stochastic dynamics
and these initial measures. In the domain-wall and zero-magnetization settings, earlier tracer arguments anticipated the
scaling exponents and limiting shapes but not the model-specific amplitudes. A corresponding current law for a homogeneous
state at nonzero magnetization had also not been worked out.

In this paper we formulate three conjectures for the time-integrated spin current. For homogeneous initial data with nonzero
magnetization, we predict a Gaussian limit on the \(t^{1/4}\) scale. For a biased domain wall, we predict that the same
tracer fluctuations are folded into the previously anticipated half-normal limit. For homogeneous initial data at zero
magnetization, we give the fully normalized M-Wright law for the XNOR process on the anticipated \(t^{1/8}\) scale,
equivalently represented by a two-sided Brownian motion evaluated at an independent Gaussian argument. In all three cases we specify the complete
proposed normalization.
To the best of our knowledge, the model-specific amplitudes, the precise formulations for Bernoulli initial states, and the moment
statements have not previously been given for the continuous-time XNOR process.

The statements are deliberately presented here as conjectures. We give the exact coordinate transformations and the
physical and probabilistic arguments supporting these conjectures, together with numerical tests.
An AI-generated candidate proof is presented in the mathematical companion \cite{PozsgayXNORProof2026}.

\subsection{The use of AI in this work}

Generative-AI systems were used extensively in the analytical and numerical work. We used ChatGPT 5.5 and 5.6 Sol. The
text of the paper was written in collaboration between the author and the AI systems.

The author selected the research problem, specified the stochastic model and observables, evaluated and edited the
generated material, and made the decision to release the manuscripts. The author carefully checked all the physical
arguments and  the mathematical derivations in this paper.

The numerical simulations
were executed in \texttt{Python}; the AI generated the simulation code, the analysis scripts, and the figures presented
in this paper.
The numerical tests reported in Section~\ref{sec:numerical-tests} support the tested conjectural limits, including the
nontrivial amplitudes and distribution shapes. The programs and the data are available on reasonable request.

The mathematical companion \cite{PozsgayXNORProof2026} was generated substantially with generative-AI assistance. The
author has reviewed its 
global structure and selected calculations, but has not completed an independent line-by-line
verification of every technical argument. Much of that material lies beyond the expertise of the author. Due to its
unverified status the companion is described as a candidate proof draft, and it should be viewed as a separate
document, which will not be submitted to peer review in its current form. We invite qualified readers to review the
mathematical manuscript 
\cite{PozsgayXNORProof2026}, either to confirm it or to point out potential mistakes in the attempted rigorous
proof. Future versions could be verified rigorously, either by a human mathematician, or by formal verification systems. 

We made the decision to release both documents, because \cite{PozsgayXNORProof2026} could become verified in a later
stage. Furthermore, we intended to
document both the capabilities and limitations of current AI-assisted mathematical research.

We stress again that all
the material that is presented in this paper was checked by the author, and only \cite{PozsgayXNORProof2026} contains
unverified computations. 

Our disclosure of AI use and assignment of human responsibility are informed by the recommendations of the
\textit{Leiden Declaration on Artificial Intelligence and Mathematics} \cite{Leiden}.

\subsection{The structure of this paper}

Section \ref{sec:model-current-initial-data} defines the model, the physical observables, and the initial measures that
are used to define the current statistics. Section \ref{sec:main-results} includes our main results: three conjectures for three
different physical initial conditions. Section \ref{sec:physical-picture} describes the strategy of the analytic computations, including
the coordinate transformations that enable us to transfer known rigorous results for the symmetric simple exclusion
process (SSEP) to this problem. This section
also explains how the mapping of physical observables remains non-trivial despite the apparent simplicity of the
coordinate space maps; these details explain the difficulty of producing a fully rigorous proof. Section
\ref{sec:numerical-tests} includes 
numerical tests of all three conjectures. We present our Conclusions in Section \ref{sec:concl}.

\section{Model, current, and initial measures}
\label{sec:model-current-initial-data}

Throughout the paper, $\1\{A\}$ denotes the indicator of the condition $A$: it equals $1$ if $A$ holds and $0$ otherwise.
Unless stated otherwise, $\PP(A)$ denotes the probability of an event $A$, and $\EE[X]$ denotes the expectation of a random variable $X$.  In both cases, the randomness includes both the initial state and the stochastic time evolution.

We now describe the XNOR hopping process, a kinetically constrained stochastic hopping model in one dimension.

Let $\eta=(\eta_j)_{j\in\ZZ}\in\{+1,-1\}^{\ZZ}$, and let
$S_j^z=\eta_j/2$ denote the physical spin component.  The dynamics is continuous in time.  Whenever four consecutive spins have the form $a\,b\,c\,a$, with $b\ne c$, the two middle spins exchange with rate one:
\begin{equation}
  \eta_j\eta_{j+1}\eta_{j+2}\eta_{j+3}=a\,b\,c\,a
  \quad\longleftrightarrow\quad
  a\,c\,b\,a,
  \qquad b\ne c.
  \label{eq:xnor-rule}
\end{equation}
The outer spins remain fixed and act as a kinetic constraint on the motion of the middle pair.

We measure the integrated spin current through the fixed bond between sites $0$ and $1$.  Each allowed exchange across this bond contributes $+1$ when a $+$ spin moves from site $0$ to site $1$, and $-1$ when it moves in the opposite direction.  Let $N_{\rightarrow}(t)$ and $N_{\leftarrow}(t)$ denote the corresponding numbers of exchanges up to time $t$.  Our current is
\begin{equation}
  J(t):=N_{\rightarrow}(t)-N_{\leftarrow}(t).
  \label{eq:current-definition}
\end{equation}
Equivalently,
\[
  J(t)=\sum_{j\ge1}\bigl(S_j^z(t)-S_j^z(0)\bigr)
  =\frac12\sum_{j\ge1}\bigl(\eta_j(t)-\eta_j(0)\bigr).
\]
Thus $J(t)$ is the net physical spin transported into the right half of the chain.  Our sign convention is that transport from left to right is positive.

We consider two initial conditions: a homogeneous product state and a domain-wall product state.  The first is the
homogeneous case. For $-1<m<1$, let $\mu_m^{\mathrm{hom}}$ be the homogeneous Bernoulli product measure
\begin{equation}
  \PP_{\mu_m^{\mathrm{hom}}}(\eta_j=+1)=\frac{1+m}{2},
  \qquad j\in\ZZ .
  \label{eq:homogeneous-measure}
\end{equation}
When the process starts from \eqref{eq:homogeneous-measure}, we write
$\Jhom(t)$ for the current \eqref{eq:current-definition}.  The first
main result below treats $0<m<1$.  The case $m=0$ is degenerate for
the leading colour-mean mechanism and is treated separately.

The second initial state is the domain-wall product state.
For $0<m<1$, let $\mu_m^{\mathrm{dw}}$ be the biased domain-wall product measure
\begin{equation}
  \PP_{\mu_m^{\mathrm{dw}}}(\eta_j=+1)=\frac{1+m}{2}\quad (j\le0),
  \qquad
  \PP_{\mu_m^{\mathrm{dw}}}(\eta_j=+1)=\frac{1-m}{2}\quad (j\ge1).
  \label{eq:domain-wall-measure}
\end{equation}
For the domain-wall initial state, we fix $m$ and suppress it in the current notation.  The notation
$\Jdw(t)$ means exactly the current \eqref{eq:current-definition} for
the stochastic XNOR dynamics started from
\eqref{eq:domain-wall-measure}.
Thus the density interface coincides with the observed current cut
between sites $0$ and $1$.

Throughout the paper, whenever the normalized magnetization parameter $m$ is used, we write
\begin{equation}
  p=\frac{1+m}{2},\qquad q=\frac{1-m}{2},
  \qquad \varphi=pq=\frac{1-m^2}{4}.
  \label{eq:pqphi}
\end{equation}
In particular, $p=q=1/2$ at zero magnetization.

\section{Main results}
\label{sec:main-results}

In this section we present the main results of this work.

First we define Gaussian and half-normal probability distributions.
For $0<m<1$, define the Gaussian variance parameter
\begin{equation}
  \sigma_m^2=
  \frac{2m^2(1-m^2)}{(3+m^2)\sqrt\pi},
  \label{eq:main-results-current-scale}
\end{equation}
and let us introduce a Gaussian variable $G_m$ with probability distribution
\begin{equation}
  G_m\sim\mathcal N(0,\sigma_m^2).
  \label{eq:main-results-physical-gaussian}
\end{equation}
Here the notation $\mathcal N(0,\sigma^2)$ denotes the normal, or Gaussian, distribution with mean zero and variance $\sigma^2$.

We also use the following half-normal notation.  For $\sigma>0$,
\begin{equation}
  \PHN(x;\sigma)=
  \frac{\sqrt2}{\sigma\sqrt\pi}
  \exp\!\left[-\frac{x^2}{2\sigma^2}\right]\Theta(x).
  \label{eq:half-normal-density}
\end{equation}
Here $\Theta$ is the Heaviside step function, with $\Theta(x)=1$ for $x\ge0$ and $\Theta(x)=0$ for $x<0$.  Equivalently, if $G\sim\mathcal N(0,\sigma^2)$, then $|G|$ has density \eqref{eq:half-normal-density}.
Let $H_m$ denote the half-normal random variable with scale $\sigma_m$:
\begin{equation}
  \PP(H_m\in\dd x)=\PHN(x;\sigma_m)\,\dd x,
  \qquad x\ge0.
  \label{eq:main-results-half-normal-variable}
\end{equation}
Equivalently, $H_m$ has the same distribution as $|G_m|$.
Thus $\sigma_m$ is the half-normal scale parameter, whereas
\begin{equation}
  \Var(H_m)=\left(1-\frac2\pi\right)\sigma_m^2.
  \label{eq:half-normal-actual-variance}
\end{equation}

Here and throughout, $\Longto$ denotes weak convergence.  When we
refer to FCS, we mean only the limiting
probability law together with the convergence of each fixed ordinary
moment stated below.  We do not claim convergence of moment generating
functions or scaled cumulant generating functions.

\subsection{Homogeneous biased product state}
\label{subsec:main-homogeneous-biased}

\begin{conjecture}[Homogeneous biased current]
\label{thm:main-summary-homogeneous}
For the process started from $\mu_m^{\mathrm{hom}}$ with $0<m<1$,
\begin{equation}
  t^{-1/4}\Jhom(t)\Longto G_m.
  \label{eq:main-results-Jhom-limit}
\end{equation}
Moreover, for every integer $r\ge1$,
\begin{equation}
 \lim_{t\to\infty}\EE[(t^{-1/4}\Jhom(t))^r]=\EE[G_m^r].
 \label{eq:main-results-Jhom-moments}
\end{equation}
\end{conjecture}

The $t^{1/4}$ scale is known from earlier work.  Singh, Ware, Vasseur, and Friedman established $z=4$ for random XNOR
circuits, while Feldmeier, Witczak-Krempa, and Knap derived the corresponding single-file tracer description and the
normalized long-time spin-correlation profile of the discrete-time circuit
\cite{SinghWareVasseurFriedman2021,FeldmeierWitczakKrempaKnap2022}.  Neither work gives the fixed-cut current law for
homogeneous biased Bernoulli data in the continuous-time  process.
To the best of our knowledge, the Gaussian fixed-cut current conjecture with the explicit \(m\)-dependent variance
coefficient \eqref{eq:main-results-current-scale} has not previously been stated for the continuous-time  process.

\subsection{Domain-wall product state}
\label{subsec:main-domain-wall}

\begin{conjecture}[Biased domain-wall current]
\label{thm:main-summary-domain-wall}
For the process started from $\mu_m^{\mathrm{dw}}$ with $0<m<1$,
\begin{equation}
  t^{-1/4}\Jdw(t)\Longto H_m.
  \label{eq:main-results-Jdw-limit}
\end{equation}
Moreover, for every integer $r\ge1$,
\begin{equation}
 \lim_{t\to\infty}\EE[(t^{-1/4}\Jdw(t))^r]=\EE[H_m^r].
 \label{eq:main-results-Jdw-moments}
\end{equation}
\end{conjecture}

The restriction $m>0$ fixes the orientation of the bias and of the
domain wall.  Global spin flip preserves the dynamics, maps
$\mu_m^{\rm hom}$ to $\mu_{-m}^{\rm hom}$, and changes the sign of the
current.  Hence the homogeneous conjecture extends to $-1<m<0$ with the
same centred Gaussian law and variance.

If the domain-wall family is defined for negative $m$ by the same formula
\eqref{eq:domain-wall-measure}, global spin flip instead gives the reflected
half-normal limit $-H_{|m|}$.

Gopalakrishnan, Morningstar, Vasseur, and Khemani derived the normalized single-interface half-normal FCS for the folded
XXZ setting and argued, with numerical support, that the same limiting shape applies to stochastic XNOR after replacing
the fluctuation scale by $t^{1/4}$ \cite{GopalakrishnanMorningstarVasseurKhemani2024}.  They did not determine the
microscopic current amplitude for stochastic XNOR.

If Conjecture~\ref{thm:main-summary-domain-wall} holds, the proposed half-normal law determines the fixed-order cumulant asymptotics.
For each fixed integer $r\ge1$,
\[
  \kappa_r[\Jdw(t)]
  =
  \kappa_r(H_m)t^{r/4}+o(t^{r/4}).
\]
Thus, when cumulants are divided by the variance scale $t^{1/2}$, every nonzero cumulant of order $r>2$ diverges as a power of time.  This is the same anomalous single-file FCS behaviour observed in earlier works
\cite{KrajnikSchmidtPasquierIlievskiProsen2022,KrajnikSchmidtPasquierProsenIlievski2024,KrajnikKlobasBertiniProsen2025,YoshimuraKrajnik2025}.

\subsection{Homogeneous zero-magnetization state}
\label{subsec:main-homogeneous-zero}

For the process started from $\mu_0^{\mathrm{hom}}$, let $X_0$ be a random variable with probability density
\begin{equation}
  p_0(x)=
  \int_0^\infty
  \frac{\dd s}{\pi a_0\sqrt{s}}
  \exp\!\left[-\frac{s^2}{2a_0^2}-\frac{x^2}{2s}\right],
  \qquad
  a_0=\frac{1}{\sqrt{6}\pi^{1/4}}.
  \label{eq:main-results-m0-density}
\end{equation}
The density in \eqref{eq:main-results-m0-density} is called the M-Wright density.  In the notation of
Ref.~\cite{FujimotoEtAl2026},
\[
  p_0(x)=P_{\mathrm{MW}}(x,a_0).
\]
Equivalently, it is a rescaled M-Wright function of order \(1/4\); the random-time Brownian representation below gives a
probabilistic realization of the same law.
The parameter $a_0$ is not the standard deviation of $X_0$.  The actual variance is
\begin{equation}
  \Var(X_0)=a_0\sqrt{\frac2\pi}.
  \label{eq:m0-actual-variance}
\end{equation}

\begin{conjecture}[Homogeneous zero-magnetization current]
\label{thm:main-summary-zero}
For the process started from $\mu_0^{\mathrm{hom}}$
\begin{equation}
  t^{-1/8}\Jhom(t)\Longto X_0,
  \label{eq:main-results-m0-limit}
\end{equation}
and, for every integer $r\ge1$,
\begin{equation}
 \lim_{t\to\infty}\EE[(t^{-1/8}\Jhom(t))^r]=\EE[X_0^r].
 \label{eq:main-results-m0-moments}
\end{equation}
\end{conjecture}

The proposed limiting variable has a useful probabilistic representation.  Let $\mathcal B=(\mathcal B_u)_{u\in\RR}$ be a two-sided standard Brownian motion, and let
\[
  G_{(0)}\sim\mathcal N\!\left(0,\frac{2}{3\sqrt\pi}\right)
\]
be independent of $\mathcal B$.  Then
\begin{equation}
  X_0\stackrel{\mathrm d}{=}\mathcal B_{-G_{(0)}/2}.
  \label{eq:main-results-m0-brownian-representation}
\end{equation}
The parenthesized subscript is deliberate: $G_{(0)}$ is the zero-magnetization specialization of the uncoloured tracer
variable introduced below, and is not the limit of $G_m$ as $m\downarrow0$; the latter is degenerate. Conditional on
$G_{(0)}$, this is a centred Gaussian variable with random variance
\[
  V_0=\frac{|G_{(0)}|}{2}.
\]
The variable $V_0$ is half-normal with scale $a_0=1/(\sqrt6\,\pi^{1/4})$.  Averaging the conditional Gaussian density over $V_0$ gives exactly the integral formula \eqref{eq:main-results-m0-density}.

This representation also makes the connection to previous FCS results explicit. The functional form of this M-Wright
density is the single-interface equilibrium Gaussian variance mixture derived by Gopalakrishnan,
Morningstar, Vasseur, and Khemani in their Eq.~(12)
\cite{GopalakrishnanMorningstarVasseurKhemani2024}. Equivalently, in the normalization used here its characteristic function is
\begin{equation}
  \EE\!\left[\exp(\mathrm{i}kX_0)\right]
  =
  \exp\!\left(\frac{a_0^2k^4}{8}\right)
  \operatorname{erfc}\!\left(\frac{a_0k^2}{2\sqrt2}\right).
  \label{eq:main-results-m0-characteristic-function}
\end{equation}
Here $\operatorname{erfc}(z)=1-\operatorname{erf}(z)$ is the complementary error function. Their periodic-boundary
expression, Eq.~(15), contains the square of the corresponding single-interface characteristic function because there are
two independent interfaces. Their transferred quantity is $\Delta M_R-\Delta M_L=2\Delta M_R$, where $M_L$ and $M_R$
denote the physical $S^z$ magnetizations of the two half-chains, whereas the
fixed-cut current $J$ in \eqref{eq:current-definition} is $\Delta M_R$; this additional convention must also be accounted for
when comparing amplitudes.

After accounting for these conventions, the general XNOR argument of
Ref.~\cite{GopalakrishnanMorningstarVasseurKhemani2024} already anticipates the $t^{1/8}$ scale and the M-Wright shape in
Conjecture~\ref{thm:main-summary-zero}. What it leaves undetermined is the amplitude for the continuous-time  process
in the present one-cut convention. The new content here is the explicit value of $a_0$.

\subsection{Comparison with standard SSEP}
\label{subsec:ssep-comparison}

To make the comparison precise, we first define SSEP and its fixed-cut current in the notation used here.  We then
compare the fixed-cut current fluctuations of SSEP with the three XNOR conjectures above, using analogous homogeneous
and domain-wall product measures.  SSEP is diffusive, with density perturbations spreading over distances of order
$t^{1/2}$.  By contrast, spin transport in the XNOR process is subdiffusive, with a spreading length of order $t^{1/4}$.
As we shall see, the different spreading exponents do not imply a uniform change in
the current-fluctuation scale; the result depends on the initial state.

In spin language, standard SSEP exchanges the nearest-neighbour spins $(\eta_j,\eta_{j+1})$ at rate one for every bond $(j,j+1)$.  Equivalently, the occupation variables
\begin{equation}
  n_j=\frac{1+\eta_j}{2}\in\{0,1\}
\end{equation}
perform nearest-neighbour exclusion dynamics.  Let $J_{\SSEP}(t)$ denote the integrated spin current across the same
fixed cut, with right jumps of the $+$ spins counted positively.  Hydrodynamic and fluctuation results for SSEP are
classical \cite{Spohn1991,KipnisLandim1999}; for the Bernoulli step measures used here the exact long-time current
generating function was computed by Derrida and Gerschenfeld \cite{DerridaGerschenfeld2009}.

For the homogeneous product state, the mean occupations on the two sides are
\[
  \bar n_L=\bar n_R=\bar n=\frac{1+m}{2}.
\]
In this equilibrium case $\EE[J_{\SSEP}(t)]=0$, and the fixed-bond central limit theorem (CLT) gives the following limit
\cite{Spohn1991,KipnisLandim1999,DerridaGerschenfeld2009}:
\[
  t^{-1/4}J_{\SSEP}(t)
  \Longto
  \mathcal N\!\left(0,\Sigma_{\SSEP,\mathrm{hom}}^2(m)\right).
\]
The limiting variance in this formula is
\begin{equation}
  \Sigma_{\SSEP,\mathrm{hom}}^2(m)
  =
  \frac{2\bar n(1-\bar n)}{\sqrt\pi}
  =
  \frac{1-m^2}{2\sqrt\pi}.
  \label{eq:ssep-homogeneous-variance}
\end{equation}
For every fixed $m>0$, the homogeneous fixed-cut currents therefore have the same leading form in the two models: both
have zero mean and Gaussian fluctuations on the $t^{1/4}$ scale.  At this level, the only difference is the variance
coefficient.  The distinction becomes qualitative at $m=0$.  Standard SSEP still has $t^{1/4}$ Gaussian current
fluctuations, with rescaled variance $1/(2\sqrt\pi)$, whereas the XNOR variance coefficient
\eqref{eq:main-results-current-scale} vanishes as $m\downarrow0$; at $m=0$ the XNOR leading scale is instead $t^{1/8}$
and the limit is the non-Gaussian variable $X_0$.

For the domain-wall product state,
\[
  \bar n_L=\frac{1+m}{2},
  \qquad
  \bar n_R=\frac{1-m}{2}.
\]
The exact SSEP current statistics \cite{DerridaGerschenfeld2009} give
\[
  \EE[J_{\SSEP}(t)]
  \sim
  m\sqrt{\frac{t}{\pi}},
\]
while the centred fluctuations obey
\begin{equation}
  \frac{J_{\SSEP}(t)-\EE[J_{\SSEP}(t)]}{t^{1/4}}
  \Longto
  \mathcal N\!\left(0,\Sigma_{\SSEP,\mathrm{dw}}^2(m)\right),
  \qquad
  \Sigma_{\SSEP,\mathrm{dw}}^2(m)
  =
  \frac{1+m^2}{2\sqrt\pi}
  -\frac{m^2}{\sqrt{2\pi}}.
\end{equation}
Thus
standard SSEP has a nonzero mean current of order $t^{1/2}$ in the domain-wall state, with smaller fluctuations around
that mean, whereas the XNOR domain-wall conjecture predicts no corresponding $t^{1/2}$ drift: its leading current is
already on the $t^{1/4}$ scale and has the proposed half-normal limit $H_m$.

\section{From XNOR dynamics to current-fluctuation laws}
\label{sec:physical-picture}

In this section we lay out the main strategy and present 
the analytic derivation supporting the conjectures.

First we explain the physical picture.  In a local move such as
\[
  ++-+\qquad\longleftrightarrow\qquad+-++,
\]
the update can be interpreted as moving the pair of equal spins by two lattice sites while retaining its sign. It might
seem counter-intuitive to focus on the pair and not on the single propagating spin (which is the negative spin above);
the idea is that the motion of the one-site excitations will be calculated based on the motion of the pairs.
The 
pairs are neither created nor 
destroyed, and their order along the chain never changes.  We may therefore mark one of these pairs near the observation
cut and follow it as a tracer.  Since the tracer cannot pass the other equal-spin pairs, it undergoes single-file motion
and typically travels a distance of order $t^{1/4}$.

The tracer is therefore a bookkeeping device whose displacement records the net number of conserved pairs crossing the
observation cut.  Each
crossing carries the sign of the corresponding equal-spin pair.  The physical spin current can therefore be viewed as a
signed sum over the pairs swept past the cut by the tracer.

The rest of this section makes this picture quantitative.  We first study only the tracer motion, temporarily ignoring
whether each equal-spin pair is $++$ or $--$, and relate this motion to a tagged particle in SSEP.  We then convert the
tracer displacement into the number of pairs that cross the fixed cut.  Finally, we reintroduce the signs carried by
these pairs to recover the physical spin current.  These steps lead to the three different fluctuation laws stated in
Section~\ref{sec:main-results}.

\begin{figure}[t]
\centering
\resizebox{\textwidth}{!}{%
\begin{tikzpicture}[
  x=0.64cm,
  y=0.85cm,
  spin/.style={
    draw=black!55,
    fill=white,
    minimum width=0.48cm,
    minimum height=0.48cm,
    inner sep=0pt,
    font=\small
  },
  bond/.style={
    draw=black!55,
    fill=white,
    minimum width=0.43cm,
    minimum height=0.43cm,
    inner sep=0pt,
    font=\small
  },
  zsite/.style={
    draw=black!65,
    fill=white,
    rounded corners=1.5pt,
    minimum width=0.66cm,
    minimum height=0.66cm,
    inner sep=0pt
  },
  ssepsite/.style={
    circle,
    draw=black!70,
    fill=white,
    minimum size=0.48cm,
    inner sep=0pt,
    font=\scriptsize
  },
  transform/.style={->, line width=0.7pt, black!65},
  tracer/.style={draw=red!75!black, fill=red!12, line width=1pt}
]
  \node[anchor=east,align=right] at (2.25,6.3)
    {{\small\bfseries initial spin chain}\\[-0.15em]$\eta_j$};
  \foreach \s [count=\k from 0] in {+,+,-,+,-,+,+,-,-,+,-,-,+,+} {
    \pgfmathsetmacro{\xpos}{3+\k}
    \node[spin] at (\xpos,6.3) {$\s$};
  }
  \node[spin,tracer] at (10,6.3) {$-$};
  \node[spin,tracer] at (11,6.3) {$-$};
  \draw[red!75!black,line width=1pt] (9.62,6.68)--(11.38,6.68);
  \node[font=\scriptsize,text=red!75!black] at (10.5,7.0) {tracer};

  \node[anchor=east,align=right] at (2.25,4.5)
    {{\small\bfseries bond variables}\\[-0.15em]$d_j$};
  \foreach \d [count=\k from 0] in {0,1,1,1,1,0,1,0,1,1,0,1,0} {
    \pgfmathsetmacro{\xpos}{3.5+\k}
    \node[bond] at (\xpos,4.5) {$\d$};
  }
  \node[bond,tracer] at (10.5,4.5) {$0$};
  \node[font=\tiny] at (3.5,4.02) {$i=-2$};
  \node[font=\tiny] at (8.5,4.02) {$i=-1$};
  \node[font=\tiny,text=red!75!black] at (10.5,4.02) {$i=0$};
  \node[font=\tiny] at (13.5,4.02) {$i=1$};
  \node[font=\tiny] at (15.5,4.02) {$i=2$};

  \node[anchor=east,align=right] at (2.25,2.4)
    {{\small\bfseries zero-range process}\\[-0.15em]gap sites};
  \draw[black!35,line width=0.6pt] (6,2.4)--(14.5,2.4);
  \foreach \xpos in {6,9.5,12,14.5} {
    \node[zsite] at (\xpos,2.4) {};
  }
  \fill[black!80] (6,2.27) circle[radius=1.8pt];
  \fill[black!80] (6,2.53) circle[radius=1.8pt];
  \fill[black!80] (12,2.4) circle[radius=1.8pt];
  \node[font=\scriptsize] at (6,1.78) {$\xi_{-2}=2$};
  \node[font=\scriptsize] at (9.5,1.78) {$\xi_{-1}=0$};
  \node[font=\scriptsize] at (12,1.78) {$\xi_{0}=1$};
  \node[font=\scriptsize] at (14.5,1.78) {$\xi_{1}=0$};
  \draw[red!75!black,line width=1.2pt] (10.5,1.95)--(10.5,2.85);
  \node[font=\scriptsize,text=red!75!black] at (10.5,3.14) {tracer bond};

  \node[anchor=east,align=right] at (2.25,0.1)
    {{\small\bfseries squeezed SSEP}\\[-0.15em]sites};
  \draw[black!35,line width=0.6pt] (4.4,0.1)--(15.25,0.1);
  \node[ssepsite] at (4.4,0.1) {$0$};
  \node[ssepsite,fill=black!80,text=white] at (5.95,0.1) {$1$};
  \node[ssepsite,fill=black!80,text=white] at (7.5,0.1) {$1$};
  \node[ssepsite] at (9.05,0.1) {$0$};
  \node[ssepsite,tracer] at (10.6,0.1) {$0$};
  \node[ssepsite,fill=black!80,text=white] at (12.15,0.1) {$1$};
  \node[ssepsite] at (13.7,0.1) {$0$};
  \node[ssepsite] at (15.25,0.1) {$0$};
  \node[font=\tiny] at (4.4,-0.43) {$i=-2$};
  \node[font=\tiny] at (9.05,-0.43) {$i=-1$};
  \node[font=\tiny,text=red!75!black] at (10.6,-0.43) {$i=0$};
  \node[font=\tiny] at (13.7,-0.43) {$i=1$};
  \node[font=\tiny] at (15.25,-0.43) {$i=2$};

  \draw[transform] (16.8,5.92)--(16.8,4.88);
  \node[anchor=west,align=left,font=\scriptsize] at (17.15,5.4)
    {compare neighbouring\\spins};
  \draw[transform] (16.8,4.08)--(16.8,2.82);
  \node[anchor=west,align=left,font=\scriptsize] at (17.15,3.45)
    {count complete $11$ pairs\\in each gap};
  \draw[transform] (16.8,1.92)--(16.8,0.52);
  \node[anchor=west,align=left,font=\scriptsize] at (17.15,1.22)
    {insert $\xi_i$ particles between\\labelled holes};
\end{tikzpicture}
}
\caption{The exact coordinate transformations for one example initial configuration.  A bond variable is zero when its two
neighbouring spins are equal.  The zero bonds are labelled consecutively by the variable $i$.  The zero-range occupation
$\xi_i$ counts the complete $11$ pairs between zero labels $i$ and $i+1$.  The parity information---whether an unpaired
$1$ remains in a gap---is stored separately in $\eps_i\in\{1,2\}$.  The squeezed SSEP contains $\xi_i$ particles
between the corresponding labelled holes.  Red
follows the same tracer through all four rows: the marked equal-spin pair, its zero bond, the bond between the adjacent
zero-range sites, and the corresponding labelled SSEP hole.}
\label{fig:coordinate-transformations}
\end{figure}
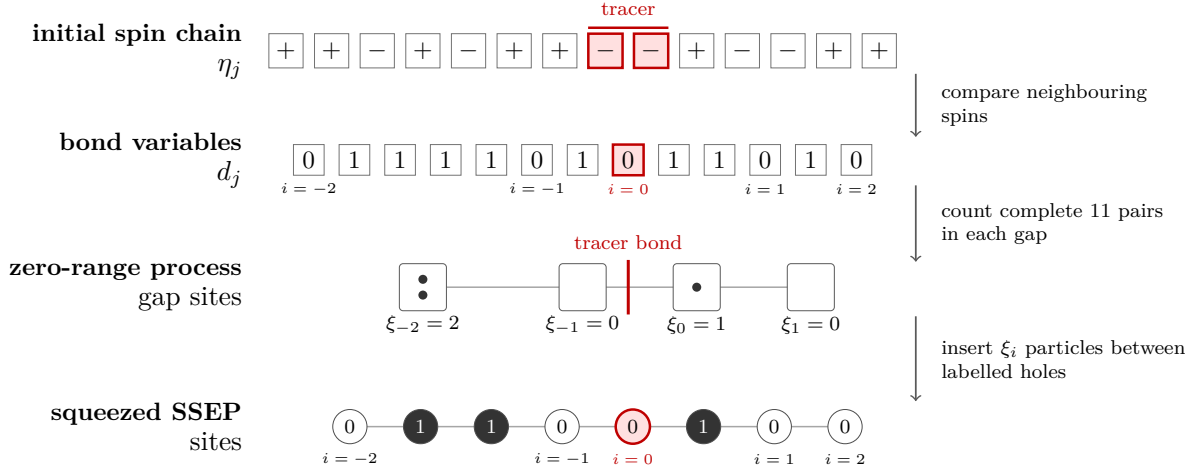

The required sequence of transformations is explained in the subsections below and
summarized in Figure~\ref{fig:coordinate-transformations}.

\subsection{Bond variables and mobile zero bonds}
\label{subsec:physical-bonds}

Introduce the bond variables
\[
  d_j=\1\{\eta_j\ne\eta_{j+1}\}.
\]
A bond with $d_j=0$ is a pair of equal spins.  We call it a zero bond, and it carries a colour, namely the common spin value $\eta_j=\eta_{j+1}\in\{+1,-1\}$.  The XNOR move becomes, in the three adjacent bond variables,
\[
  011\longleftrightarrow110 .
\]
Thus the local move does not create or destroy zero bonds.  It moves one zero bond by two bond sites through a neighbouring pair of ones.  Consequently the parity of each zero-bond position is conserved, and zero bonds never overtake one another.

The only bond-variable move that transports spin across the physical cut between sites $0$ and $1$ is a zero-bond jump
from bond site $-1$ to bond site $1$, or the reverse jump.  Both sites are odd, and zero bonds preserve their parity
under the dynamics.  We will call zero bonds on odd bond sites active,
and the tracer that we introduce below to represent the fixed-cut current will be chosen from the active zero bonds.
When an active zero bond of colour $a$
jumps from $-1$ to $1$, the spin current increases by $a$; a jump from $1$ to $-1$ contributes $-a$.  Thus the spin
current is the signed colour sum of the active zero bonds that cross the fixed cut.

If we write a zero bond as a vacancy symbol $\circ$ and a one bond as a material symbol $\bullet$, then
\[
  011\longleftrightarrow110
  \qquad\text{becomes}\qquad
  \circ\,\bullet\,\bullet
  \longleftrightarrow
  \bullet\,\bullet\,\circ .
\]
This is exactly the kinematic motion of a length-two hard rod through a vacancy. 
This hard-rod picture also connects the present stochastic process with two closely related quantum constructions.  In bond variables, the folded XXZ chain has the same allowed local move $011\leftrightarrow110$ \cite{PozsgayEtAl2021}.  The hard-rod deformation of quantum spin chains replaces mobile spins by hard rods of length $\ell$, producing analogous kinematic rules \cite{PozsgayGomborHutsalyuk2021}.  Here we use the inverse operation: every mobile pair $11$ is squeezed into a point particle.  This is carried out in two steps below.

\subsection{Gap variables and zero-range motion}
\label{subsec:physical-hard-rods}

Label the zero bonds consecutively by $i\in\ZZ$, assigning label $0$ to a chosen reference bond.  Between two consecutive zero bonds the bond word contains only ones.  Thus a gap has the form
\[
  0\,1\,1\,\cdots\,1\,0 .
\]
The mobile objects in such a gap are the complete pairs $11$.  If the number of intervening ones is odd, one unpaired $1$ remains.  This unpaired one is not mobile by itself; it is frozen sector data.  Formally, for consecutive zero labels $i$ and $i+1$ we write the distance as
\[
  R_i=\eps_i+2\xi_i,
  \qquad
  \eps_i\in\{1,2\},\quad \xi_i\in\NN_0.
\]
The variable $\xi_i$ counts the number of mobile $11$ rods in the gap, while $\eps_i$ records the frozen parity residue.  The zero-bond colours are another frozen layer: they move with the zero labels, but they do not affect which uncoloured moves are allowed.

We can now forget the detailed bond word for a moment and look only at the collection $(\xi_i)_{i\in\ZZ}$.  This collection can be seen as a particle configuration: site $i$ is the gap between zero labels $i$ and $i+1$, and it contains $\xi_i$ particles, namely the mobile $11$ rods in that gap.  In this auxiliary particle system, one site $i$ may contain any number of particles.

The XNOR dynamics induces a closed dynamics on these occupation numbers.  When a pair $11$ crosses a zero bond, it leaves one gap and enters the neighbouring gap.  In the occupation variables this is just one particle moving from site $i$ to site $i-1$ or from site $i$ to site $i+1$.  Thus $(\xi_i)$ may be viewed as a Markov dynamics on its own.

This dynamics has the zero-range property: the rate at which a particle leaves site $i$ depends only on $\xi_i$, the occupation of the departure site, and not on the occupations of the neighbouring sites.  More precisely, if $\xi_i>0$, one particle can jump from site $i$ to either neighbouring site, while if $\xi_i=0$ no particle can leave.  Since each allowed edge move has rate one, $(\xi_i)$ evolves as a constant-rate zero-range process.

We now determine the distribution of the gap variables induced by homogeneous Bernoulli initial data and show that the
resulting distribution is an equilibrium measure for the zero-range dynamics.
Let $c_i\in\{+1,-1\}$ denote the colour carried by zero bond $i$.  Consider first the gaps viewed from a specified zero
bond.  Suppose that $c_i=+1$.  Starting from
this bond, the spins alternate until the next zero bond is reached.  The two zero bonds therefore have the same colour
when their distance is odd, corresponding to $\eps_i=1$, and opposite colours when their distance is even, corresponding
to $\eps_i=2$.  If the gap contains $k$ mobile $11$ pairs, the two
possibilities have probabilities
\begin{align*}
 \PP(\xi_i=k,\eps_i=1,c_{i+1}=+1\mid c_i=+1)&=p(pq)^k,\\
 \PP(\xi_i=k,\eps_i=2,c_{i+1}=-1\mid c_i=+1)&=q^2(pq)^k.
\end{align*}
The corresponding formulas for a zero bond of colour $-1$ follow by interchanging $p$ and $q$.  Adding the two
possibilities gives
\[
 \PP(\xi_i=k\mid c_i=+1)=(p+q^2)(pq)^k=(1-pq)(pq)^k,
\]
and the same result follows for $c_i=-1$, since $q+p^2=1-pq$.  Therefore every gap occupation has the geometric
distribution
\begin{equation}
  \PP(\xi_i=k)=\nu_\varphi(k)=(1-\varphi)\varphi^k,
  \qquad k\in\NN_0,
  \qquad
  \rho:=\EE[\xi_i]=\frac{\varphi}{1-\varphi},
  \label{eq:phys-geometric-gap-law}
\end{equation}
where $\varphi=pq$.  The common factor $(pq)^k$ in these expressions is the only dependence on $k$; after normalization, the probabilities for the colour and
parity of the next zero bond do not depend on the gap occupation.  Moreover, once the next zero bond has been reached,
the spins farther along the chain have not yet been inspected and remain independent Bernoulli variables.  Repeating the
same calculation gap by gap, in both directions, shows that the variables $(\xi_i)$ are independent and are also
independent of the complete frozen colour and parity chain $(c_i,\eps_i)$.
The frozen variables $(c_i,\eps_i)$ themselves form a Markov chain and are generally correlated from one gap to the
next; their explicit distribution will not be needed below.

The product geometric distribution is stationary under the zero-range dynamics.  Indeed, a jump changes two adjacent
occupations from $(a,b)$ to $(a-1,b+1)$.  Their product weight is proportional to $\varphi^{a+b}$ both before and after
the jump, and the forward and reverse rates are both one.  Thus the mobile particles and the frozen colour information
separate both in the dynamics and in the equilibrium statistics.

The fact that the distributions of the gap variables are stationary also follows from the stationarity of the original
homogeneous measure: the gap variables are simply just new variables to represent the same physical configurations, and
if the original measure is in equilibrium, then the distributions of all new variables are also stationary.

This bulk equilibrium law is relevant to all three initial measures.  For either homogeneous initial measure it applies
throughout the bulk.  In the domain-wall initial measure, the Bernoulli parameters $p$ and $q$ are interchanged between
the two tails, but the gap parameter $\varphi=pq$ is unchanged.  The two tails therefore have the same uncoloured
equilibrium gap statistics.

There is one qualification common to all three initial measures.  The geometric law above describes gaps viewed from a
chosen zero bond, whereas the physical current is measured at a fixed spatial cut, which is more likely to lie in a long
gap.  Consequently, the nearby gap variables do not initially have exactly the equilibrium distribution above.  In
Subsection~\ref{subsec:physical-domain-wall-defect} we explain how a local replacement makes this difference negligible
on the fluctuation scales.

\subsection{The squeeze to SSEP}
\label{subsec:physical-ssep}

The constant-rate zero-range process has a second exact representation as SSEP.  Write SSEP holes as $0$'s and SSEP particles as $1$'s.  Put one labelled SSEP hole for each zero label, and put $\xi_i$ SSEP particles between the holes corresponding to labels $i$ and $i+1$:
\[
  \cdots 0_i\,1^{\xi_i}\,0_{i+1}\,1^{\xi_{i+1}}\,0_{i+2}\cdots .
\]
Here $1^{\xi_i}$ means a string of $\xi_i$ consecutive $1$'s, and the subscript on a $0$ is only the label of the
corresponding original zero bond.  Then a zero-range particle leaving a gap is simply the SSEP particle at the edge of
that gap exchanging with the adjacent hole.  The squeezed exclusion lattice is not the original spin lattice.  It is the
coordinate line obtained after collapsing each mobile rod $11$ to one point and deleting the frozen parity residues from
the moving coordinate system. 

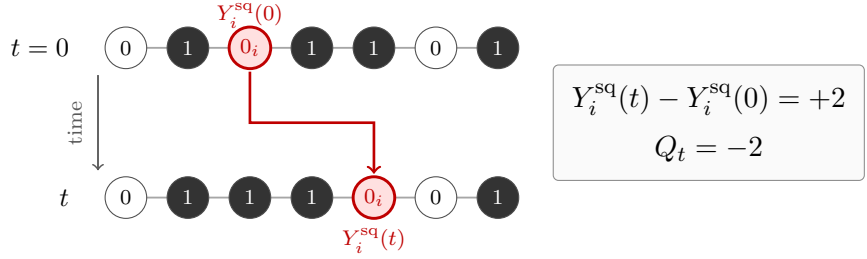
\begin{figure}[t]
\centering
\begin{tikzpicture}[
  x=0.82cm,
  y=0.9cm,
  ssephole/.style={
    circle,
    draw=black!70,
    fill=white,
    minimum size=0.55cm,
    inner sep=0pt,
    font=\scriptsize
  },
  ssepparticle/.style={
    ssephole,
    fill=black!80,
    text=white
  },
  taggedhole/.style={
    ssephole,
    draw=red!75!black,
    fill=red!12,
    text=red!75!black,
    line width=1.1pt
  }
]
  \node[anchor=east,font=\small] at (1.25,2.3) {$t=0$};
  \draw[black!35,line width=0.7pt] (2,2.3)--(8,2.3);
  \node[ssephole] at (2,2.3) {$0$};
  \node[ssepparticle] at (3,2.3) {$1$};
  \node[taggedhole] at (4,2.3) {$0_i$};
  \node[ssepparticle] at (5,2.3) {$1$};
  \node[ssepparticle] at (6,2.3) {$1$};
  \node[ssephole] at (7,2.3) {$0$};
  \node[ssepparticle] at (8,2.3) {$1$};
  \node[font=\scriptsize,text=red!75!black] at (4,2.78) {$Y_i^{\rm sq}(0)$};

  \node[anchor=east,font=\small] at (1.25,0.1) {$t$};
  \draw[black!35,line width=0.7pt] (2,0.1)--(8,0.1);
  \node[ssephole] at (2,0.1) {$0$};
  \node[ssepparticle] at (3,0.1) {$1$};
  \node[ssepparticle] at (4,0.1) {$1$};
  \node[ssepparticle] at (5,0.1) {$1$};
  \node[taggedhole] at (6,0.1) {$0_i$};
  \node[ssephole] at (7,0.1) {$0$};
  \node[ssepparticle] at (8,0.1) {$1$};
  \node[font=\scriptsize,text=red!75!black] at (6,-0.53) {$Y_i^{\rm sq}(t)$};

  \draw[->,black!65,line width=0.7pt] (1.55,1.9)--(1.55,0.5);
  \node[rotate=90,font=\scriptsize,text=black!65] at (1.18,1.2) {time};

  \draw[->,red!75!black,line width=1pt]
    (4,1.95)--(4,1.2)--(6,1.2)--(6,0.45);

  \node[
    draw=black!45,
    rounded corners=2pt,
    fill=black!2,
    align=center,
    inner sep=7pt
  ] at (11.4,1.2) {
    $\displaystyle Y_i^{\rm sq}(t)-Y_i^{\rm sq}(0)=+2$\\[0.45em]
    $\displaystyle Q_t=-2$
  };
\end{tikzpicture}
\caption{Definition of the auxiliary tracer current in the squeezed SSEP.  The red hole carries the fixed label $i$ and
is followed throughout the evolution.  Its position at time $t$ is $Y_i^{\rm sq}(t)$, and $Q_t$ is defined as minus its
net displacement.  In the example shown, the tagged hole moves two sites to the right, so its displacement is $+2$ and
$Q_t=-2$.}
\label{fig:auxiliary-tracer-current}
\end{figure}

In this representation the labelled zero bond becomes a tagged SSEP hole.  Choose such a tagged hole, say the one with label $i$, and let $Y_i^{\rm sq}(t)$ be its position in the squeezed SSEP lattice.  We define
\[
  Q_t=-\bigl(Y_i^{\rm sq}(t)-Y_i^{\rm sq}(0)\bigr)
\]
for minus its tagged-hole displacement, and we suppress the dependence on $i$.  We call $Q_t$ the auxiliary tracer current.
Figure~\ref{fig:auxiliary-tracer-current} illustrates this definition and its sign convention.  Equivalently, it is the signed
zero-range current through the bond between the two gaps adjacent to the chosen zero label: a zero-range particle crossing
from gap $i-1$ to gap $i$ counts as $+1$, and the reverse crossing counts as $-1$.

Let $Z_i(t)$ denote the position of
zero bond $i$ on the original bond lattice.  For an active zero bond, define its position on the active lattice by
\[
  A_i(t)=\frac{Z_i(t)+1}{2}.
\]
The corresponding displacements then satisfy
\begin{equation}
  Z_i(t)-Z_i(0)=-2Q_t,
  \qquad
  A_i(t)-A_i(0)=-Q_t,
  \label{eq:phys-exact-coordinate-current}
\end{equation}
and this follows from the no-overtaking property of SSEP.

So far we have not yet specified the index $i$, and $Q_t$ clearly depends on $i$ (although we suppressed the notation). 
We specify $i$ in the next subsection, in order to make a connection to the physical current.

However, we can highlight the main use of this definition already at this point:
the auxiliary tracer current  equals the displacement of a tagged hole in one-dimensional
equilibrium SSEP, up to the sign convention above.  When the gap occupations are independently distributed according to
\eqref{eq:phys-geometric-gap-law}, we can use the classical tagged-particle result
\cite{Arratia1983,PeligradSethuraman2008} 
\begin{equation}
  t^{-1/4}Q_t\Longto G_Q,
  \qquad
  G_Q\sim\mathcal N\!\left(0,\frac{2\rho}{\sqrt\pi}\right).
  \label{eq:phys-tagged-hole-limit}
\end{equation}
For $m=0$, we denote this limiting variable by $G_{(0)}$; thus $G_{(0)}$ is the zero-magnetization specialization of $G_Q$.
The local replacement described in Subsection~\ref{subsec:physical-domain-wall-defect} transfers this limit from a tracer initialized in equilibrium to the tracer selected at the fixed physical cut.  These $t^{1/4}$ single-file fluctuations are the first pillar of the reduction; the same tracer viewpoint also appears in the physics literature on constrained and charged single-file dynamics \cite{SinghWareVasseurFriedman2021,FeldmeierWitczakKrempaKnap2022,KrajnikSchmidtPasquierProsenIlievski2024}.

Knowing the auxiliary tracer current $Q_t$ is not yet enough to determine the spin current.  Two further steps are
needed.  First, the tagged-hole displacement has to be converted into a current across a 
fixed spin cut; this uses the no-overtaking order of zero labels and requires us to count the active zeros between the
tracer and the cut.  Second, the spin colours have to be restored by adding the colours of the zero bonds that cross.  This
is where the homogeneous biased, domain-wall biased, and homogeneous $m=0$ limits separate.

\subsection{From tracer displacement to current across the fixed cut}
\label{subsec:physical-fixed-cut}

The SSEP tagged-particle limit \eqref{eq:phys-tagged-hole-limit} describes the displacement of one labelled zero bond,
whereas the physical observable counts crossings of a fixed spatial cut.  This subsection relates these two quantities.
Because zero bonds cannot overtake one another, the labels that change sides of the cut are precisely those in the
interval swept out by the tagged zero.

In a homogeneous region, the probability that an active site contains a zero bond is
\begin{equation}
  \delta=\PP(d_{2y-1}=0)=p^2+q^2=\frac{1+m^2}{2}.
  \label{eq:phys-active-density}
\end{equation}
This formula applies to all three initial measures.  In either homogeneous initial measure, $\delta$ is the bulk
active-zero density throughout the chain.  In the domain-wall initial measure, changing the magnetization from $m$ to
$-m$ merely interchanges $p$ and $q$, so both tails again have the same density $\delta$.  Thus the same leading
conversion between tracer displacement and active-zero current applies whichever direction the tracer moves.

The central asymptotic relation required is
\begin{equation}
  N_t=-\delta Q_t+o_{\PP}(t^{1/4}).
  \label{eq:phys-fixed-cut-target}
\end{equation}
Here $o_{\PP}(t^{1/4})$ denotes a random correction that is negligible on the $t^{1/4}$ scale: after division by
$t^{1/4}$, it converges to zero in probability.
We use $O_{\PP}(t^\alpha)$ for a random quantity that remains tight after division by $t^\alpha$.

In the remainder of this subsection we give the physical derivation and identify the estimates that would establish this relation rigorously.
The argument has two steps.  First, we use preservation of the zero-label order to express the fixed-cut current as an
exact count over the interval between the cut and a tagged zero.  Second, we replace this count by the active-zero
density times the tagged-zero displacement and show that the resulting density fluctuation is subleading. 

At time $t=0$, we now choose the tracer introduced above to be the active zero nearest to the left of the cut, including
active site $0$.  We assign this zero bond label $0$ and follow it throughout the evolution.  In the notation introduced
above, its active-lattice position is $A_0(t)$.  We put
\begin{equation}
  B=A_0(0)\le0,
  \qquad
  \Delta A_t=A_0(t)-B=-Q_t.
  \label{eq:phys-tag-position}
\end{equation}
Let $\zeta_y(t)$ be the indicator that active site $y$ is occupied by an active zero.  In terms of the original bond variables,
\[
  \zeta_y(t)=\1\{d_{2y-1}(t)=0\}=1-d_{2y-1}(t).
\]
Let $N_t$ be the signed active-zero current across the active bond $(0,1)$, positive from left to right.  No overtaking makes the conversion from the tag to the fixed cut an exact order-statistic identity:
\begin{equation}
  N_t=
  \begin{cases}
    \displaystyle\sum_{y=1}^{A_0(t)}\zeta_y(t),&A_0(t)\ge1,\\[0.7em]
    0,&A_0(t)=0,\\[0.3em]
    \displaystyle-\sum_{y=A_0(t)+1}^{0}\zeta_y(t),&A_0(t)\le-1.
  \end{cases}
  \label{eq:phys-order-statistic}
\end{equation}
Indeed, the labels which have changed side are exactly the active zeros occupying the interval swept by the tagged zero.

This identity does not assume that the tracer leaves the surrounding configuration unchanged.  The active zeros move as
well, and their positions at time $t$ are correlated with the tracer position.  The previous computations about the
equilibrium distributions enter when we estimate
the number of active zeros in the final interval.  For either homogeneous initial measure, the Bernoulli spin measure is
stationary under the XNOR dynamics, consistently with the stationary product gap measure found in
Subsection~\ref{subsec:physical-hard-rods}.  Thus an interval of length $L$ at any fixed time typically contains
$\delta L$ active zeros, with fluctuations of order $L^{1/2}$.  Because the endpoint is the random tracer position, this
estimate must be controlled simultaneously for all intervals on the relevant $t^{1/4}$ scale.  In the domain-wall case,
the two homogeneous tails have the same density $\delta$, while the finite interfacial region is treated separately.

We therefore write the exact count as its mean-density contribution plus a fluctuation $E_t$.  If only the mean-density
contribution were present, the oriented count in \eqref{eq:phys-order-statistic} would be $\delta A_0(t)$.  We define
$E_t$ as the difference between the actual count and this mean value.  Since $A_0(t)=B+\Delta A_t$, the exact relation is
\begin{equation}
  N_t=\delta\Delta A_t+\delta B+E_t,
  \label{eq:phys-exact-density-decomposition}
\end{equation}
where
\begin{equation}
  E_t=
  \begin{cases}
    \displaystyle\sum_{y=1}^{A_0(t)}(\zeta_y(t)-\delta),&A_0(t)\ge1,\\[0.7em]
    0,&A_0(t)=0,\\[0.3em]
    \displaystyle-\sum_{y=A_0(t)+1}^{0}(\zeta_y(t)-\delta),&A_0(t)\le-1.
  \end{cases}
  \label{eq:phys-density-count-error}
\end{equation}
$E_t$ is simply the local density fluctuation in the contiguous lattice interval between the fixed cut and the tagged zero.  Its
purpose is to keep track of the error in replacing the actual number of active zeros by density times distance.  The
endpoint $A_0(t)$ is itself random and is determined by the same evolving configuration as the field $\zeta_y(t)$.  We
therefore cannot directly use a fluctuation estimate for an interval with a fixed endpoint.  Instead, the candidate proof
develops a uniform estimate over all endpoints within distances of order $t^{1/4}$ from the cut, which is the distance
scale explored by the tracer.  This estimate can then be applied to the actual random endpoint $A_0(t)$.

The initial offset $B$ has an exponential tail and is therefore microscopic.  For either homogeneous initial
measure---the biased case or the zero-magnetization case---the number of active zeros in an interval of length $L$
fluctuates around $\delta L$ on the scale $L^{1/2}$.  Since the tagged zero explores distances of order $t^{1/4}$,
applying this square-root estimate uniformly over the possible tracer endpoints gives the nested scale
\[
  |E_t|=O_{\PP}(t^{1/8}).
\]
In the domain-wall case, the candidate proof develops the analogous estimate on the leading $t^{1/4}$ scale by
decomposing a window into its two homogeneous tails and one finite moving seam, without assuming independence between
the seam location and the tails.  Together, these estimates give, for all three initial measures,
\begin{equation}
  N_t=\delta\Delta A_t+o_{\PP}(t^{1/4})
  =-\delta Q_t+o_{\PP}(t^{1/4}).
  \label{eq:phys-active-current-reduction}
\end{equation}
This completes the reduction from the fixed-cut active-zero current to the auxiliary tracer current.  In the next
subsection we restore the zero-bond colours and obtain the physical spin current.

\subsection{Colour decoration and the three limits}
\label{subsec:physical-colours}

We now restore the zero-bond colours and consider the three initial measures in turn.  The active-zero current $N_t$
counts how many active zero labels cross the cut, but the spin current is their oriented colour sum.  After cancelling
recrossings, no overtaking implies that the net current is carried by a consecutive block of $|N_t|$ zero-bond labels
next to the cut.  Thus restoring the colours amounts to applying a
law of large numbers or a central limit theorem to the sum of the first $|N_t|$ entries in the appropriate initial colour
sequence.

First consider the homogeneous biased initial measure.  The conditional colour mean of an active zero bond is the same
on both sides of the cut.  We denote this mean by $\vartheta_m$ and now compute it.  A zero bond has colour $+1$ with
weight $p^2$ and colour $-1$ with weight $q^2$, and hence
\[
  \vartheta_m=\EE[\eta_j\mid d_j=0]
  =\frac{p^2-q^2}{p^2+q^2}
  =\frac{2m}{1+m^2}.
\]
Since the density of active zero bonds is $\delta=(1+m^2)/2$, the spin charge per active-coordinate site swept by the tracer is
\[
  \delta\vartheta_m=m .
\]
This identity is natural: $p^2-q^2=m$, while mixed-spin pairs do not contribute an active zero bond.  Thus the mean
charge carried by active zero bonds per active site equals the physical spin per two-site active cell.
Since $|N_t|$ is typically of order $t^{1/4}$, the fluctuation of the colour sum around its mean is only of order
$t^{1/8}$.  This gives
\[
  \Jhom(t)=\vartheta_mN_t+O_{\PP}(t^{1/8})
  =-mQ_t+o_{\PP}(t^{1/4}),
\]
which is the proposed leading relation and gives the Gaussian limit stated in Conjecture~\ref{thm:main-summary-homogeneous}.

For the domain-wall initial measure, the equality of the active-zero densities on the two sides gives
\[
  |N_t|=\delta|Q_t|+o_{\PP}(t^{1/4})
\]
with the same factor $\delta$, whichever direction the tracer moves.  The mean zero-bond colour, however, has opposite
signs on the two sides of the cut.  A right-to-left crossing also carries the opposite orientation sign.  These two sign
changes combine so that the leading contribution is always positive relative to the size of the tracer displacement:
\[
  \Jdw(t)=\vartheta_m|N_t|+O_{\PP}(t^{1/8})
  =m|Q_t|+o_{\PP}(t^{1/4}).
\]
This mechanism folds the Gaussian auxiliary tracer-current prediction into the half-normal limit stated in
Conjecture~\ref{thm:main-summary-domain-wall}.

Finally, consider the homogeneous zero-magnetization initial measure.  At time $0$, the pairs
\[
  \bigl(\eta_{2y-1}(0),\zeta_y(0)\bigr)
\]
are independent for different $y$, since they involve disjoint spin pairs, and each of their four possible values occurs
with probability $1/4$.  When $\zeta_y(0)=1$, $\eta_{2y-1}(0)$ is the colour of the corresponding zero bond.  Thus the
active-zero occupancies and colours are independent fair variables.  The physical current therefore has a nested random
structure: the tracer motion in the squeezed SSEP lattice determines a random block of zero-bond labels, and the current
is the sum of the random colours carried by that block.

We now express this structure in formulas.  If $S=(S_n)_{n\in\ZZ}$ denotes the two-sided simple symmetric random walk
formed by listing the active-zero colours outwards from the cut, then the spin current satisfies the exact identity
\begin{equation}
  \Jhom(t)=S_{N_t}
  \qquad(m=0).
  \label{eq:phys-m0-exact-random-sum}
\end{equation}
Under the usual diffusive scaling, the simple symmetric random walk converges to a two-sided Brownian motion:
\begin{equation}
  \lambda^{-1/2}S_{\lfloor\lambda u\rfloor}\Longto\mathcal B_u
  \qquad(\lambda\to\infty).
  \label{eq:phys-colour-walk-brownian-limit}
\end{equation}
In particular, the two-sided walk at index $n\in\ZZ$ typically has size $\sqrt{|n|}$.  Since $|N_t|$ is of order $t^{1/4}$, the random sum $S_{N_t}$ is therefore of order $t^{1/8}$.

The local-equilibrium mismatch at the fixed cut, already noted in
Subsection~\ref{subsec:physical-hard-rods}, is present for all three initial measures and will be treated in the next
subsection.  For the zero-magnetization random-time limit, however, it creates an additional issue: the colour walk $S$
and the auxiliary tracer current $Q_t$ are not exactly independent at finite time.  The comparison construction described
there replaces only the finite block of gap occupations around the cut, leaving the zero-bond colours and parity residues
unchanged.  It produces an equilibrium comparison current $\bar Q_t$, independent of the frozen colour environment, such
that
\[
  Q_t-\bar Q_t=O_{\PP}(1).
\]
This difference is microscopic and disappears after rescaling, allowing the limiting Brownian colour walk and Gaussian
tracer current to be taken independent.

At $m=0$ the active-zero density is $\delta=1/2$, so \eqref{eq:phys-active-current-reduction}, the auxiliary tracer-current limit
\eqref{eq:phys-tagged-hole-limit}, and the random-walk-to-Brownian limit
\eqref{eq:phys-colour-walk-brownian-limit} give
\begin{equation}
  t^{-1/4}N_t\Longto-\frac{G_{(0)}}{2},
  \qquad
  t^{-1/8}\Jhom(t)\Longto\mathcal B_{-G_{(0)}/2}.
  \label{eq:phys-m0-brownian-limit}
\end{equation}
Conditional on $G_{(0)}$, the second limiting variable is Gaussian with variance $V_0=|G_{(0)}|/2$.  Since $V_0$ is half-normal with scale $a_0$, averaging this conditional Gaussian density gives precisely $p_0(x)$ in \eqref{eq:main-results-m0-density}.  This connects the microscopic random-walk description directly to the explicit M-Wright distribution in Conjecture~\ref{thm:main-summary-zero}.

\subsection{Local equilibrium correction at the fixed cut}
\label{subsec:physical-domain-wall-defect}

The SSEP tagged-particle limit \eqref{eq:phys-tagged-hole-limit} assumes that
the tagged hole is surrounded by an equilibrium gap configuration.  The tracer
relevant to the physical XNOR current, however, is the active zero nearest to
the left of the fixed cut.  This cut-based selection biases the nearby gap
statistics: by construction, no active zero lies between the tracer and the
cut, and longer gaps are more likely to contain the cut.

To isolate the affected region, label the nearest active zero to the left of
the cut by $0$ and the nearest active zero to the right by $\tau$.  These need
not be consecutive zero labels, because inactive zero bonds may lie between
them.  The region between them corresponds to the finite block of gap
occupations
\[
  \xi_0,\ldots,\xi_{\tau-1}.
\]
This block is typically small.  On either homogeneous side, an active site
contains a zero bond with probability
\[
  \delta=p^2+q^2>0,
\]
so the distance from the cut to either bracketing active zero has an
exponentially decaying tail, and a large central block is exponentially
unlikely.

The domain-wall case has one additional feature: the same central block also
contains the density interface.  Outside the two bracketing active zeros, the
spins have their original homogeneous Bernoulli distributions, with densities
$p$ and $q=1-p$ on the two sides.  As noted in
Subsection~\ref{subsec:physical-hard-rods}, interchanging $p$ and $q$ leaves the
equilibrium gap parameter unchanged:
\[
  \varphi=pq.
\]
Thus, for all three initial measures, only the finite central block differs
from the equilibrium gap distribution.

We remedy this situation with a comparison construction that lets us apply the
equilibrium SSEP theorem while keeping track of the error caused by modifying
the gaps near the cut.  The idea is to construct an auxiliary gap field and
evolve it together with the original field using the same random updates.  The
main steps are described below, while further details of the proposed
construction and estimates are presented in the mathematical companion
\cite{PozsgayXNORProof2026}.

We introduce an auxiliary initial gap field $\bar\xi(0)$.  Inside the central
block, the occupations $\bar\xi_i(0)$, $i=0,\ldots,\tau-1$, are sampled
independently from the equilibrium geometric law $\nu_\varphi$ in
\eqref{eq:phys-geometric-gap-law}.  Outside the block we set
\[
  \bar\xi_i(0)=\xi_i(0).
\]
The original field is not changed.  The complete auxiliary field has the
equilibrium product distribution required for the tagged-particle limit, while
the actual and auxiliary fields differ only near the cut.

The replacement retains every zero label, its colour, and its parity residue.
Since
\[
  R_i=\eps_i+2\xi_i,
\]
changing $\xi_i$ changes a gap length only by an even number and does not
insert or remove zero bonds.  The actual and auxiliary configurations
therefore live on the same gap-index lattice and can be compared bond by bond.

We couple their evolutions using the same family of  Poisson clocks.
For every gap site $i$, we use separate clocks for jumps $i\to i+1$ and
$i\to i-1$.  When either clock rings, one particle jumps in each configuration
for which the departure site $i$ is occupied.  Thus both configurations jump if
both departure sites are occupied, while only one jumps if the other departure
site is empty.  Each configuration separately follows the correct zero-range
dynamics.

To measure their difference at the same time, define
\[
  \Delta_i(t)=\xi_i(t)-\bar\xi_i(t),
\]
and let
\begin{equation}
  D=\sum_{i\in\ZZ}|\Delta_i(0)|.
  \label{eq:phys-defect-size}
\end{equation}
Here $D$ is the initial total occupation mismatch, not a spatial distance.
Because the length of the resampled block has an exponential tail and the
geometric gap occupations also have exponential tails, $D$ has an
exponentially decaying tail.

Under the coupled evolution, a unit of occupation mismatch can move from one
gap to a neighbouring gap.  If mismatches of opposite sign meet, they can
cancel.  No additional mismatch is created.  Consequently,
\begin{equation}
  \sum_{i\in\ZZ}|\Delta_i(t)|\le D
  \qquad\text{for all }t\ge0.
  \label{eq:phys-defect-stability}
\end{equation}
Thus the distance between the actual and auxiliary configurations, compared
at the same time $t$, cannot increase.

For each bond $b=(i-1,i)$ of the shared gap lattice, let $Q_b(t)$ and
$\bar Q_b(t)$ denote the currents in the actual and auxiliary evolutions,
respectively.  Their difference is the change in the total mismatch to the
right of that bond:
\[
  Q_b(t)-\bar Q_b(t)
  =\sum_{j\ge i}\bigl[\Delta_j(t)-\Delta_j(0)\bigr].
\]
It follows that
\[
  |Q_b(t)-\bar Q_b(t)|\le2D
  \qquad\text{for all }t\ge0.
\]
This estimate holds separately for every bond $b$, and in particular for the
bond associated with the tagged zero.

The difference between the auxiliary tracer current and the equilibrium
comparison tracer current is therefore bounded uniformly in time by the random
variable $2D$, which has an exponential tail.  This correction disappears on both
the $t^{1/4}$ and $t^{1/8}$ fluctuation scales.  The coupling acts only on the
uncoloured gap occupations; the zero-bond colours remain attached to their
labels throughout.

\subsection{Constants, scales, and where approximation enters}
\label{subsec:physical-constants-ledger}

The variance coefficients proposed in the main conjectures follow from the chain of reductions as follows.  From
\[
  \varphi=pq=\frac{1-m^2}{4}
\]
the mean zero-range occupation is
\begin{equation}
  \rho=\frac{\varphi}{1-\varphi}
  =\frac{1-m^2}{3+m^2}.
  \label{eq:phys-rho-m}
\end{equation}
The tagged-hole limit \eqref{eq:phys-tagged-hole-limit} gives the uncoloured variance parameter $2\rho/\sqrt\pi$.  Multiplication by the physical colour coefficient $\delta\vartheta_m=m$ therefore gives
\begin{equation}
  m^2\frac{2\rho}{\sqrt\pi}
  =\frac{2m^2(1-m^2)}{(3+m^2)\sqrt\pi},
  \label{eq:phys-variance-flow}
\end{equation}
which is exactly \eqref{eq:main-results-current-scale}.  The domain wall applies the absolute-value map to the same Gaussian scale.  At $m=0$, the mean colour of an active zero bond is zero, so the contribution proportional to $Q_t$ is absent.  The current is instead governed by the fluctuations of a zero-mean colour sum over a block whose typical length is $t^{1/4}$.  Its typical size is therefore $t^{1/8}$.

It is useful to summarize the logical status of the principal relations.  The bond transformation, the gap decomposition, the zero-range/SSEP squeeze, the coordinate identities \eqref{eq:phys-exact-coordinate-current}, the order statistic \eqref{eq:phys-order-statistic}, and the zero-magnetization random-sum identity \eqref{eq:phys-m0-exact-random-sum} are  exact.  Besides the finite local-equilibrium replacement described above, the two scale-dependent counting approximations replace the number of active zeros in the swept interval by density times distance and the colour sum over the crossing labels by mean colour times their number:
\[
  N_t=-\delta Q_t+o_{\PP}(t^{1/4}),
  \qquad
  \Jhom(t)=-mQ_t+o_{\PP}(t^{1/4}),
  \qquad
  \Jdw(t)=m|Q_t|+o_{\PP}(t^{1/4}).
\]
Both errors are ordinary counting fluctuations over contiguous lattice intervals beginning at the cut.  The candidate
proof seeks to establish these bounds uniformly over endpoints on the relevant scale, which is the key point needed to
handle the random tracer endpoint without an independence assumption.

Finally, weak convergence of the SSEP tracer does not by itself imply
convergence of moments.  The candidate proof also addresses the additional
estimates needed to obtain convergence of every fixed moment
\cite{PozsgayXNORProof2026}.

\section{Numerical tests of the current-fluctuation conjectures}
\label{sec:numerical-tests}

We numerically test Conjectures~\ref{thm:main-summary-homogeneous},
\ref{thm:main-summary-domain-wall}, and \ref{thm:main-summary-zero} directly in
their three respective initial measures.

We simulate the microscopic continuous-time  XNOR process directly on
the finite interval of sites $-L,\ldots,L+1$ and measure the signed current
through the fixed bond between sites $0$ and $1$.  We use open boundary
conditions: only four-site windows contained in this interval are updated,
with no periodic wrap-around or boundary injection.  We sample the process
using a rejection-free Gillespie algorithm \cite{Gillespie1976} and assess
finite-volume effects by comparing simulations at different values of $L$.

All comparisons below are parameter-free: the scale and shape of each limiting
law are fixed by the corresponding conjecture, and no location, scale, or
shape parameter is fitted.  To compare the complete distributions, we use the
first Wasserstein distance
\begin{equation}
 W_1(\mu,\nu)
 =\int_{\mathbb R}|F_\mu(x)-F_\nu(x)|\,\mathrm dx,
 \label{eq:numerics-wasserstein-distance}
\end{equation}
which is the area between their cumulative distribution functions (CDFs).  Smaller
values indicate closer agreement at the tested time.

Statistical uncertainties are estimated by resampling the independent Monte
Carlo trajectories with replacement; quoted interval estimates contain the
central 95\% of the resulting values.  Parenthesized and $\pm$ uncertainties
attached to sample means denote one Monte Carlo standard error over independent
trajectories.  Sample sizes are given with the corresponding results.

\subsection{Homogeneous biased current}
\label{sec:numerical-homogeneous-biased}

For Conjecture~\ref{thm:main-summary-homogeneous}, the initial spins are
independent with the same probability $(1+m)/2$ of $+1$ at every site, as in
\eqref{eq:homogeneous-measure}.  To match the domain-wall test below, the three
main ensembles used $L=512$, observation times
\[
  t=256, 1024, 4096, 16384,
\]
magnetizations $m=0.5,0.7,0.8$, and $10{,}000$, $20{,}000$, and $10{,}000$
trajectories, respectively.  Together they contain approximately
$1.2769\times10^{11}$ accepted exchanges.

Put
\begin{equation}
  X_t^{\mathrm{hom}}=t^{-1/4}\Jhom(t).
  \label{eq:numerics-scaled-homogeneous-current}
\end{equation}
Reflection symmetry gives a centred, symmetric finite-time distribution.  The conjecture further predicts
\begin{equation}
 \EE[(X_t^{\mathrm{hom}})^2]\longrightarrow\sigma_m^2,
 \qquad
 \EE[(X_t^{\mathrm{hom}})^4]\longrightarrow3\sigma_m^4,
 \label{eq:numerics-homogeneous-moment-targets}
\end{equation}
together with convergence of the full law to $G_m$.  No empirical centring or
scale fitting is applied.  The variance, fourth moment, and complete distribution
therefore test the conjectured scale and Gaussian shape.

\begin{figure}[t]
\centering
\includegraphics[width=\linewidth]{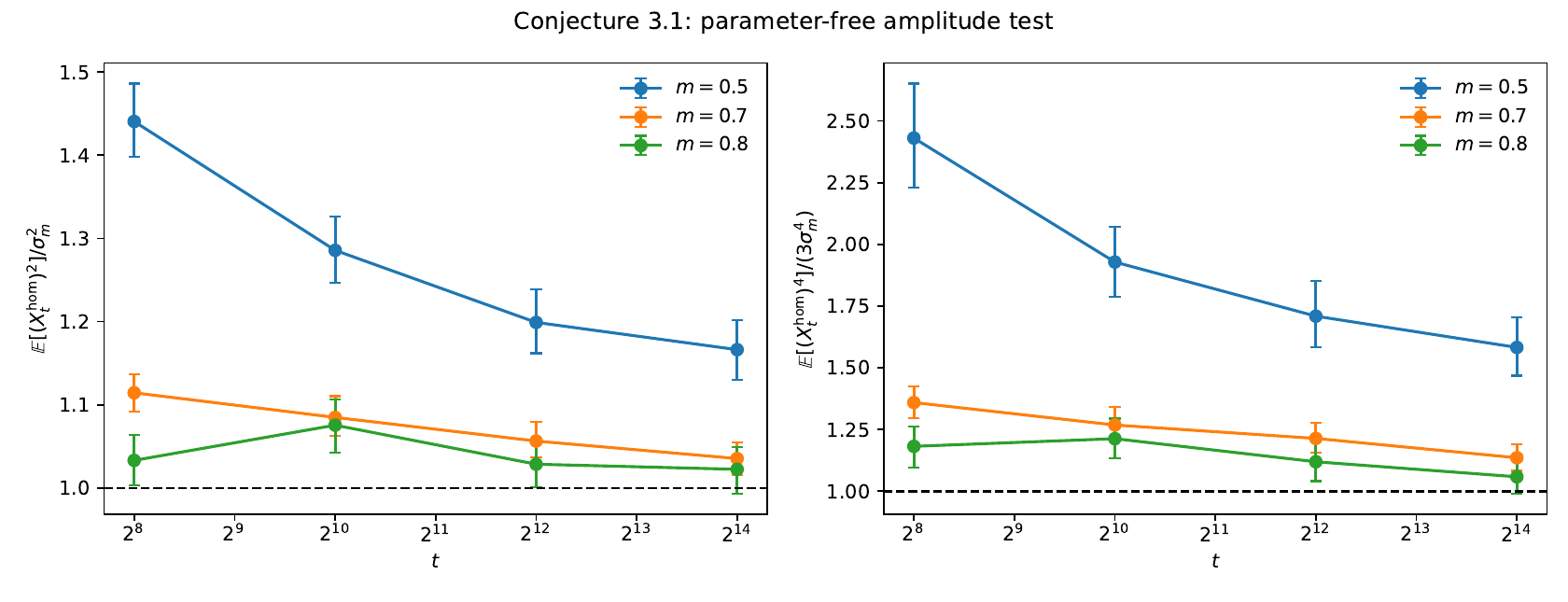}
\caption{Second- and fourth-moment tests for the homogeneous
biased problem.  The panels show
$\EE[(X_t^{\mathrm{hom}})^2]/\sigma_m^2$ and
$\EE[(X_t^{\mathrm{hom}})^4]/(3\sigma_m^4)$.  Error bars are 95\%
uncertainty intervals obtained by trajectory resampling;
the dashed lines are the predictions of
Conjecture~\ref{thm:main-summary-homogeneous}.}
\label{fig:numerics-homogeneous-moment-ratios}
\end{figure}

The results at the largest common time are collected in
Table~\ref{tab:numerics-homogeneous-largest-time}.  Parentheses after the means
are Monte Carlo standard errors over independent trajectories.  The
corresponding 95\% uncertainty intervals for the second and fourth moments are
shown in Figure~\ref{fig:numerics-homogeneous-moment-ratios}.

\begin{table}[!ht]
\centering
\small
\setlength{\tabcolsep}{3.2pt}
\begin{tabular}{c r c cc cc c}
\hline
$m$ & samples & $\EE[X_t^{\mathrm{hom}}]$ & $\EE[(X_t^{\mathrm{hom}})^2]$ & prediction & $\EE[(X_t^{\mathrm{hom}})^4]$ & prediction & $W_1$ \\
\hline
$0.5$ & $10{,}000$ & $0.00418(276)$ & $0.07593$ & $0.06510$ & $0.02012$ & $0.01271$ & $0.02791$ \\
$0.7$ & $20{,}000$ & $0.00047(205)$ & $0.08367$ & $0.08080$ & $0.02224$ & $0.01958$ & $0.02266$ \\
$0.8$ & $10{,}000$ & $0.00013(270)$ & $0.07303$ & $0.07142$ & $0.01620$ & $0.01530$ & $0.02227$ \\
\hline
\end{tabular}
\caption{Test of the homogeneous biased current at the largest common time
($t=16384$).
The first Wasserstein distance compares the empirical law of
$X_t^{\mathrm{hom}}$ with the centred Gaussian law of variance $\sigma_m^2$.
The displayed currents are not empirically centred.}
\label{tab:numerics-homogeneous-largest-time}
\end{table}

The first Wasserstein distances are collected in
Table~\ref{tab:numerics-homogeneous-wasserstein-time-series}; they decrease at
every successive observation time.
\begin{table}[H]
\centering
\small
\begin{tabular}{c cccc}
\hline
$m$ & $t=256$ & $t=1024$ & $t=4096$ & $t=16384$ \\
\hline
$0.5$ & $0.07061$ & $0.05068$ & $0.03749$ & $0.02791$ \\
$0.7$ & $0.06319$ & $0.04493$ & $0.03203$ & $0.02266$ \\
$0.8$ & $0.06322$ & $0.04454$ & $0.03152$ & $0.02227$ \\
\hline
\end{tabular}
\caption{First Wasserstein distance between the empirical homogeneous-current
distribution and the Gaussian law predicted by
Conjecture~\ref{thm:main-summary-homogeneous}, at four observation times.}
\label{tab:numerics-homogeneous-wasserstein-time-series}
\end{table}

\begin{figure}[t]
\centering
\includegraphics[width=0.8\linewidth]{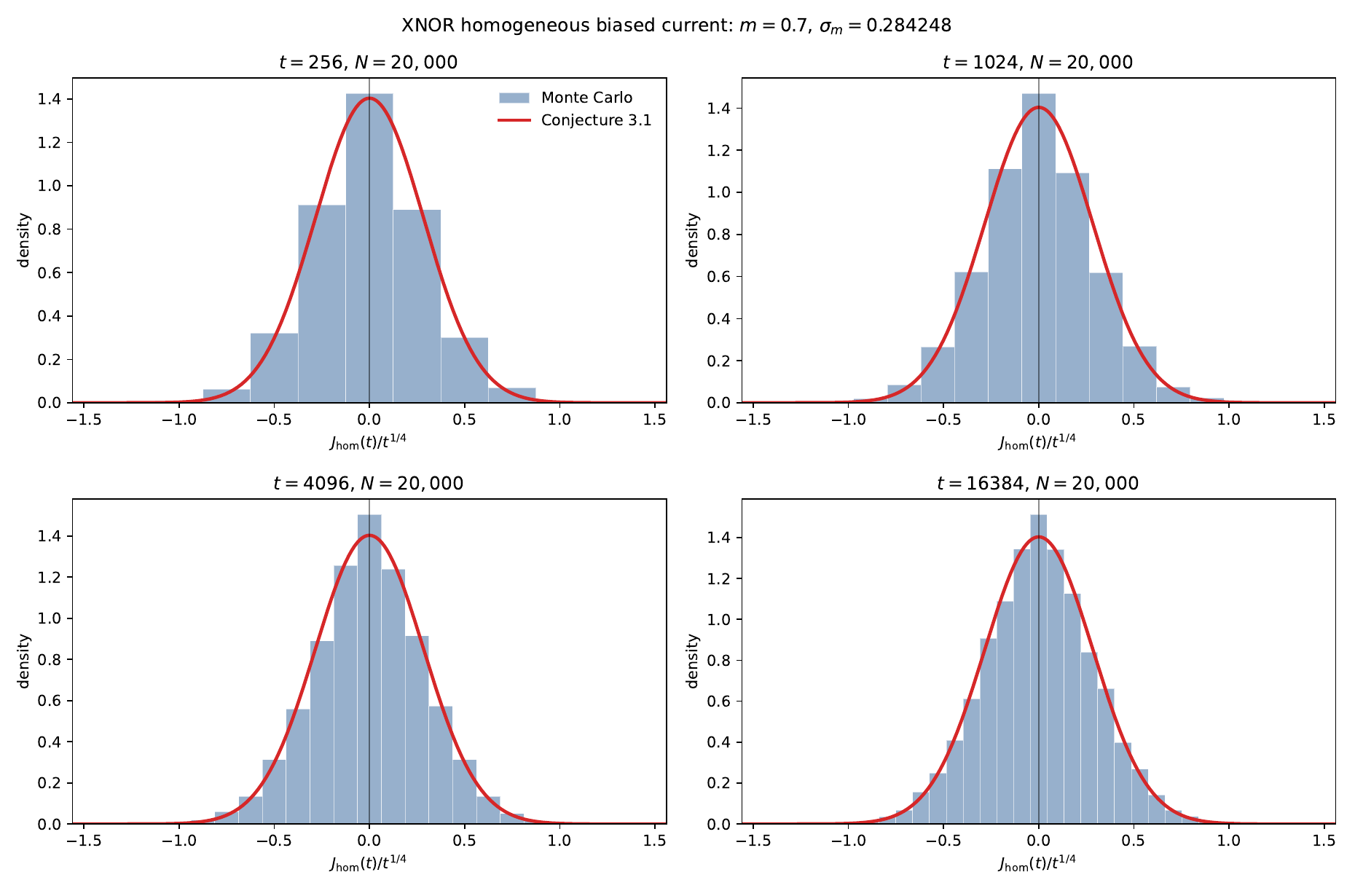}
\caption{Distribution of $\Jhom(t)/t^{1/4}$ for the homogeneous biased state
at $m=0.7$, using $L=512$ and $20{,}000$ trajectories.  Histograms are the
Monte Carlo data and the solid curve is the centred Gaussian density of
variance $\sigma_{0.7}^2$ predicted by
Conjecture~\ref{thm:main-summary-homogeneous}; no parameter is fitted.}
\label{fig:numerics-homogeneous-distributions}
\end{figure}

Figure~\ref{fig:numerics-homogeneous-distributions} shows the four-time density
comparison for the $m=0.7$ main ensemble.  The narrowing lattice spacing and
the approach to the unfitted Gaussian curve are both visible.

The second- and fourth-moment ratios over all four times are shown in
Figure~\ref{fig:numerics-homogeneous-moment-ratios}, where residual drift
remains visible in the scaled amplitudes.

We also performed an independent longer diagnostic at $m=0.7$, with $L=1024$,
$5{,}000$ trajectories, and final time $t=65536$.  It gives
\begin{equation}
 \begin{aligned}
 \EE[X_t^{\mathrm{hom}}]&=-0.00264\pm0.00406,
 &\EE[(X_t^{\mathrm{hom}})^2]&=0.08232,\\
 \EE[(X_t^{\mathrm{hom}})^4]&=0.02095,
 &W_1&=0.01618.
 \end{aligned}
 \label{eq:numerics-homogeneous-long-diagnostic}
\end{equation}
The 95\% uncertainty intervals for the second and fourth moments contain the
predictions.  At the overlapping time $t=16384$, its second moment agrees with
that of the main $L=512$ ensemble within uncertainty.

The same run follows the tagged active zero and directly tests the signed
microscopic reduction $\Jhom(t)=-mQ_t+o_{\PP}(t^{1/4})$.  At the largest time,
\begin{equation}
 \EE[(Q_t/t^{1/4})^2]=0.16271,
 \qquad
 \EE[(-mQ_t/t^{1/4})^2]=0.07973,
 \label{eq:numerics-homogeneous-tagged-diagnostic}
\end{equation}
compared with the predictions $0.16489$ and $\sigma_{0.7}^2=0.08080$.
Moreover, the root-mean-square total residual in $\Jhom(t)+mQ_t$, divided by
$t^{1/4}$, decreases from $0.14076$ at $t=1024$ to $0.07477$ at $t=65536$;
the active-zero density and colour residuals decrease as well.

The largest remaining correction is at $m=0.5$: its final second moment is
$16.6\%$ above the prediction and its fourth moment is $58.2\%$ above the
prediction.  Both ratios decrease over the tested grid, but the accessible
times cannot exclude a persistent correction.  At $m=0.7$ the amplitude correction is
smaller and the longer independent run is closer to the target; at $m=0.8$
both displayed target moments lie inside their uncertainty intervals.  Taken
together, the distributional distances, moment trends, and
signed auxiliary tracer-current reduction support Conjecture~\ref{thm:main-summary-homogeneous}
over the accessible ranges, with slow finite-time convergence at smaller bias
as the principal limitation.

\subsection{Domain-wall current}
\label{sec:numerical-domain-wall}

For Conjecture~\ref{thm:main-summary-domain-wall}, the initial state is the
Bernoulli domain wall defined in \eqref{eq:domain-wall-measure}.  The three main
ensembles had $L=512$, observation times
\[
  t=256, 1024, 4096, 16384,
\]
and magnetizations $m=0.5,0.7,0.8$.  We generated $10{,}000$, $20{,}000$, and $10{,}000$ trajectories, respectively, amounting in total to approximately $1.2760\times10^{11}$ accepted exchanges.

Put
\begin{equation}
  X_t^{\mathrm{dw}}=t^{-1/4}\Jdw(t).
  \label{eq:numerics-scaled-domain-wall-current}
\end{equation}
The predictions tested numerically are
\begin{equation}
  \EE[X_t^{\mathrm{dw}}]\longrightarrow \sigma_m\sqrt{\frac{2}{\pi}},
  \qquad
  \EE[(X_t^{\mathrm{dw}})^2]\longrightarrow \sigma_m^2,
  \label{eq:numerics-domain-wall-moment-targets}
\end{equation}
with $\sigma_m$ fixed by \eqref{eq:main-results-current-scale}, and convergence of the full distribution to
\eqref{eq:half-normal-density}.
The target scale tends to zero as either $m\downarrow0$ or $m\uparrow1$.
These limits are non-uniform: at $m=0$ the leading current changes to the
$t^{1/8}$ law tested below, while at $m=1$ the product state is frozen.

\begin{figure}[t]
\centering
\includegraphics[width=\linewidth]{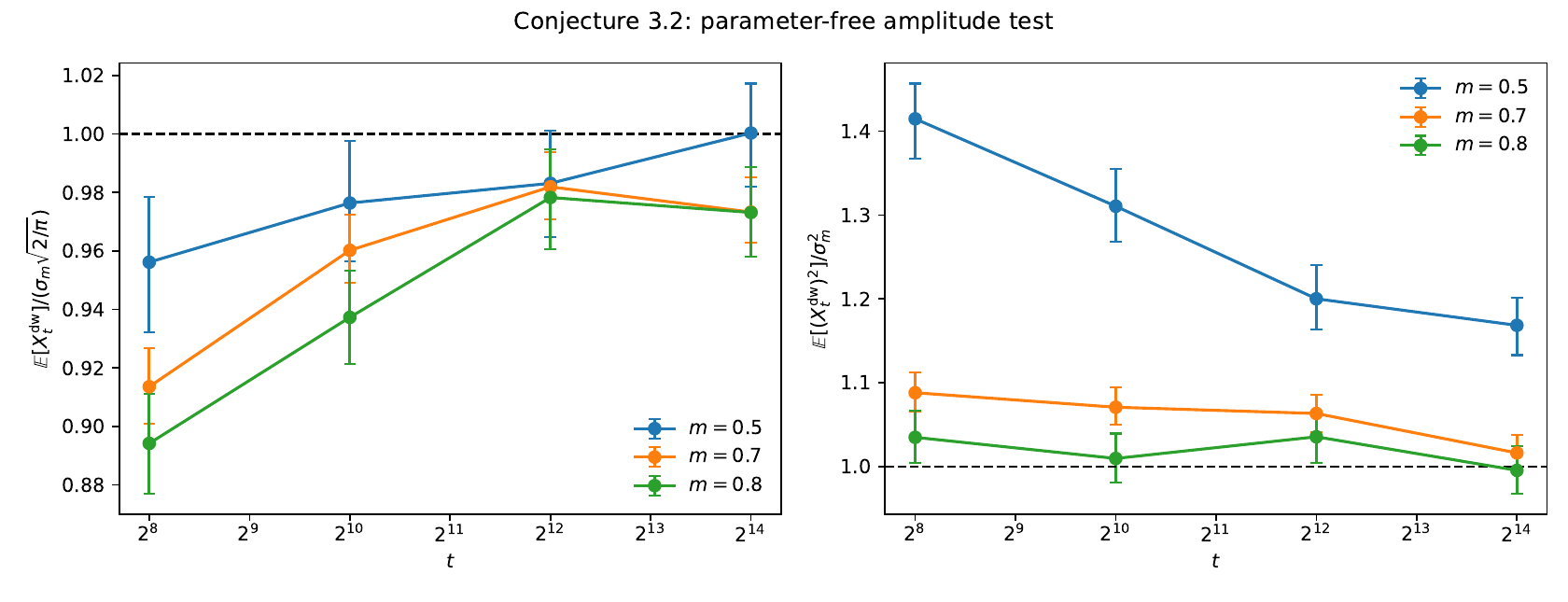}
\caption{First- and second-moment tests for the domain-wall problem with $m>0$.  The left panel shows $\EE[X_t^{\mathrm{dw}}]/(\sigma_m\sqrt{2/\pi})$ and the right panel shows $\EE[(X_t^{\mathrm{dw}})^2]/\sigma_m^2$.  Error bars are 95\% uncertainty intervals obtained by trajectory resampling; the dashed lines are the predictions of Conjecture~\ref{thm:main-summary-domain-wall}.}
\label{fig:numerics-domain-wall-moment-ratios}
\end{figure}

Table~\ref{tab:numerics-domain-wall-largest-time} gives the results at the
largest common time.  Parentheses after the empirical means are Monte Carlo
standard errors over independent trajectories.  The corresponding 95\%
uncertainty intervals for the second moments are shown in
Figure~\ref{fig:numerics-domain-wall-moment-ratios}.

\begin{table}[!ht]
\centering
\small
\setlength{\tabcolsep}{3.5pt}
\begin{tabular}{c r cc cc cc}
\hline
$m$ & samples & $\EE[X_t^{\mathrm{dw}}]$ & prediction & $\EE[(X_t^{\mathrm{dw}})^2]$ & prediction & $W_1$ & $\PP(\Jdw(t)<0)$ \\
\hline
$0.5$ & $10{,}000$ & $0.20364(186)$ & $0.20358$ & $0.07606$ & $0.06510$ & $0.03177$ & $2.890\%$ \\
$0.7$ & $20{,}000$ & $0.22074(129)$ & $0.22680$ & $0.08210$ & $0.08080$ & $0.02379$ & $0.885\%$ \\
$0.8$ & $10{,}000$ & $0.20751(167)$ & $0.21323$ & $0.07109$ & $0.07142$ & $0.02274$ & $0.260\%$ \\
\hline
\end{tabular}
\caption{Test of the domain-wall current at the largest common time
($t=16384$).  The first Wasserstein distance $W_1$ compares the empirical law of $X_t^{\mathrm{dw}}$ with the half-normal law of scale $\sigma_m$.  The last column measures the residual finite-time weight on the side forbidden by the limiting distribution.}
\label{tab:numerics-domain-wall-largest-time}
\end{table}

\begin{figure}[t]
\centering
\includegraphics[width=0.8\linewidth]{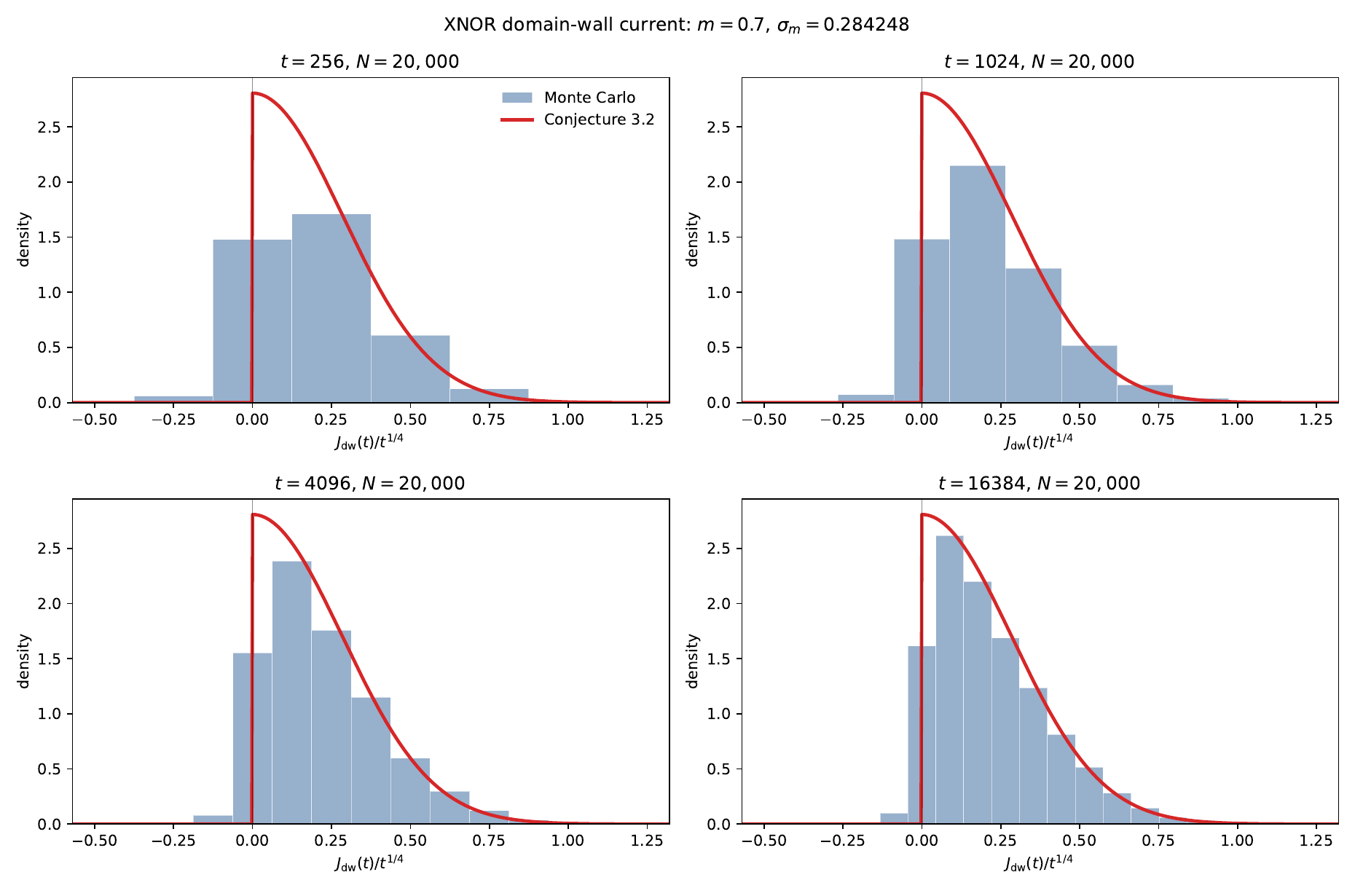}
\caption{Distribution of $\Jdw(t)/t^{1/4}$ for the $m=0.7$ domain wall at
four selected times, using $L=512$ and $20{,}000$ trajectories.  Histograms
are the Monte Carlo data and the solid curve is the half-normal density of
scale $\sigma_{0.7}$ predicted by
Conjecture~\ref{thm:main-summary-domain-wall}; no parameter is fitted.}
\label{fig:numerics-domain-wall-distributions}
\end{figure}

The first Wasserstein distances are collected in
Table~\ref{tab:numerics-domain-wall-wasserstein-time-series}; over the four
simulated observation times, they decrease for every magnetization.
\begin{table}[H]
\centering
\small
\begin{tabular}{c cccc}
\hline
$m$ & $t=256$ & $t=1024$ & $t=4096$ & $t=16384$ \\
\hline
$0.5$ & $0.08177$ & $0.06053$ & $0.04274$ & $0.03177$ \\
$0.7$ & $0.06745$ & $0.04743$ & $0.03373$ & $0.02379$ \\
$0.8$ & $0.06494$ & $0.04560$ & $0.03237$ & $0.02274$ \\
\hline
\end{tabular}
\caption{First Wasserstein distance between the empirical domain-wall current
distribution and the half-normal law predicted by
Conjecture~\ref{thm:main-summary-domain-wall}, at four observation times.}
\label{tab:numerics-domain-wall-wasserstein-time-series}
\end{table}
The finite-time distributions retain a small negative-current tail.  The histograms at all four observation times for
$m=0.7$ are shown in
Figure~\ref{fig:numerics-domain-wall-distributions}.

The moment ratios over all four times are displayed in
Figure~\ref{fig:numerics-domain-wall-moment-ratios}, where residual drift
remains visible in the scaled amplitudes.

We also performed a longer diagnostic at $m=0.7$ with $L=1024$, $5{,}000$ trajectories, and final time $t=65536$.  At
the final time it gives
\begin{equation}
  \EE[X_t^{\mathrm{dw}}]=0.22405,
  \qquad
  \EE[(X_t^{\mathrm{dw}})^2]=0.08120,
  \qquad
  W_1=0.0164,
  \label{eq:numerics-domain-wall-long-diagnostic}
\end{equation}
compared with the predictions $0.22680$ and $0.08080$.  In this run we followed the tagged active zero and measured the
intermediate currents in the reduction of Section~\ref{sec:physical-picture}.  The scaled auxiliary tracer-current second
moment was $0.16192$, compared with its Gaussian prediction $0.16489$.  Moreover, the root-mean-square total reduction
residual in $\Jdw(t)-m|Q_t|$, divided by $t^{1/4}$, decreased from $0.12763$ at $t=1024$ to $0.07297$ at $t=65536$; the
density and colour residuals also decreased.  At the overlapping time $t=16384$, the $L=1024$ run gave mean $0.21986$
and second moment $0.08151$, close to $0.22074$ and $0.08210$ for the larger-sample $L=512$ ensemble.  This supplies a
useful finite-volume check and directly tests the auxiliary tracer-current mechanism behind the conjecture.

The clearest remaining finite-time correction occurs at $m=0.5$.  Its mean at $t=16384$ agrees with the predicted value, but its second moment is still $16.8\%$ above $\sigma_{0.5}^2$, and $2.89\%$ of its currents remain negative.  The observed decrease of the first Wasserstein distance and shrinking negative tail over this time grid are consistent with slower convergence at smaller bias.  A substantially longer $m=0.5$ run would be the most useful further test.

Overall, the numerical results are consistent with
Conjecture~\ref{thm:main-summary-domain-wall} over the accessible time and
volume ranges.

\subsection{Homogeneous zero-magnetization current}
\label{sec:numerical-zero-magnetization}

For Conjecture~\ref{thm:main-summary-zero}, every site is initially chosen
independently with equal probabilities for $+1$ and $-1$.  The main ensemble
used $L=512$, observation times
\[
  t=256, 1024, 4096, 16384, 65536,
\]
and $10{,}000$ independent trajectories.  It contained approximately
$1.6757\times10^{11}$ accepted exchanges.  We recorded the auxiliary tracer
current and the active-zero current in every trajectory.

Put
\begin{equation}
  X_t^{(0)}=t^{-1/8}\Jhom(t).
  \label{eq:numerics-scaled-zero-current}
\end{equation}
The proposed limiting law tested here is the M-Wright distribution \(X_0\) in
Conjecture~\ref{thm:main-summary-zero}, represented by
\(X_0=\mathcal B_{-G_{(0)}/2}\) and hence by a Gaussian variance mixture.  In particular,
\begin{equation}
 \EE|X_0|=0.3632661012,\qquad
 \EE[X_0^2]=0.2446678852,\qquad
 \EE[X_0^4]=0.2820947918,
 \label{eq:numerics-zero-moment-targets}
\end{equation}
and its standardized kurtosis is $3\pi/2$.  The density in
\eqref{eq:main-results-m0-density} and all these amplitudes supply the
comparison.

Table~\ref{tab:numerics-zero-time-series} summarizes the observed time series.
The numerical first Wasserstein distance decreases over the five simulated
times.  Since $\Jhom(t)$ is integer-valued, $X_t^{(0)}$ lies on a lattice with
spacing $t^{-1/8}$, which is still $1/4$ at the largest time.  Because
$p_0(0)>0$, the prominent central bin is an expected lattice effect rather
than evidence for a point mass in the limit.

\begin{table}[H]
\centering
\small
\setlength{\tabcolsep}{4pt}
\begin{tabular}{r cccc}
\hline
$t$ & $\EE|X_t^{(0)}|$ & $\EE[(X_t^{(0)})^2]$ & $\EE[(X_t^{(0)})^4]$ & $W_1$ \\
\hline
$256$   & $0.33090$ & $0.23010$ & $0.19095$ & $0.12680$ \\
$1024$  & $0.34506$ & $0.23814$ & $0.22390$ & $0.10585$ \\
$4096$  & $0.34857$ & $0.23684$ & $0.22918$ & $0.08951$ \\
$16384$ & $0.35632$ & $0.24746$ & $0.27074$ & $0.07420$ \\
$65536$ & $0.35315$ & $0.24096$ & $0.25851$ & $0.06268$ \\
limit   & $0.36327$ & $0.24467$ & $0.28209$ & $0$       \\
\hline
\end{tabular}
\caption{Results for the homogeneous problem at $m=0$, using the main $L=512$
ensemble of $10{,}000$ trajectories.  The table shows the moments of the rescaled
current $X_t^{(0)}$ and its first Wasserstein distance $W_1$ from the distribution in
Conjecture~\ref{thm:main-summary-zero}.}
\label{tab:numerics-zero-time-series}
\end{table}

At $t=65536$, the 95\% uncertainty intervals for the second and fourth moments
contain the predictions, while the interval for the absolute first moment
remains slightly below its limiting value.  The sample mean
$0.00035\pm0.00491$ and the conditional positive-current fraction $0.5019$
are consistent with the predicted symmetry.
The standardized kurtosis is $4.452$, compared with $3\pi/2=4.712$.

\begin{figure}[H]
\centering
\includegraphics[width=0.8\linewidth]{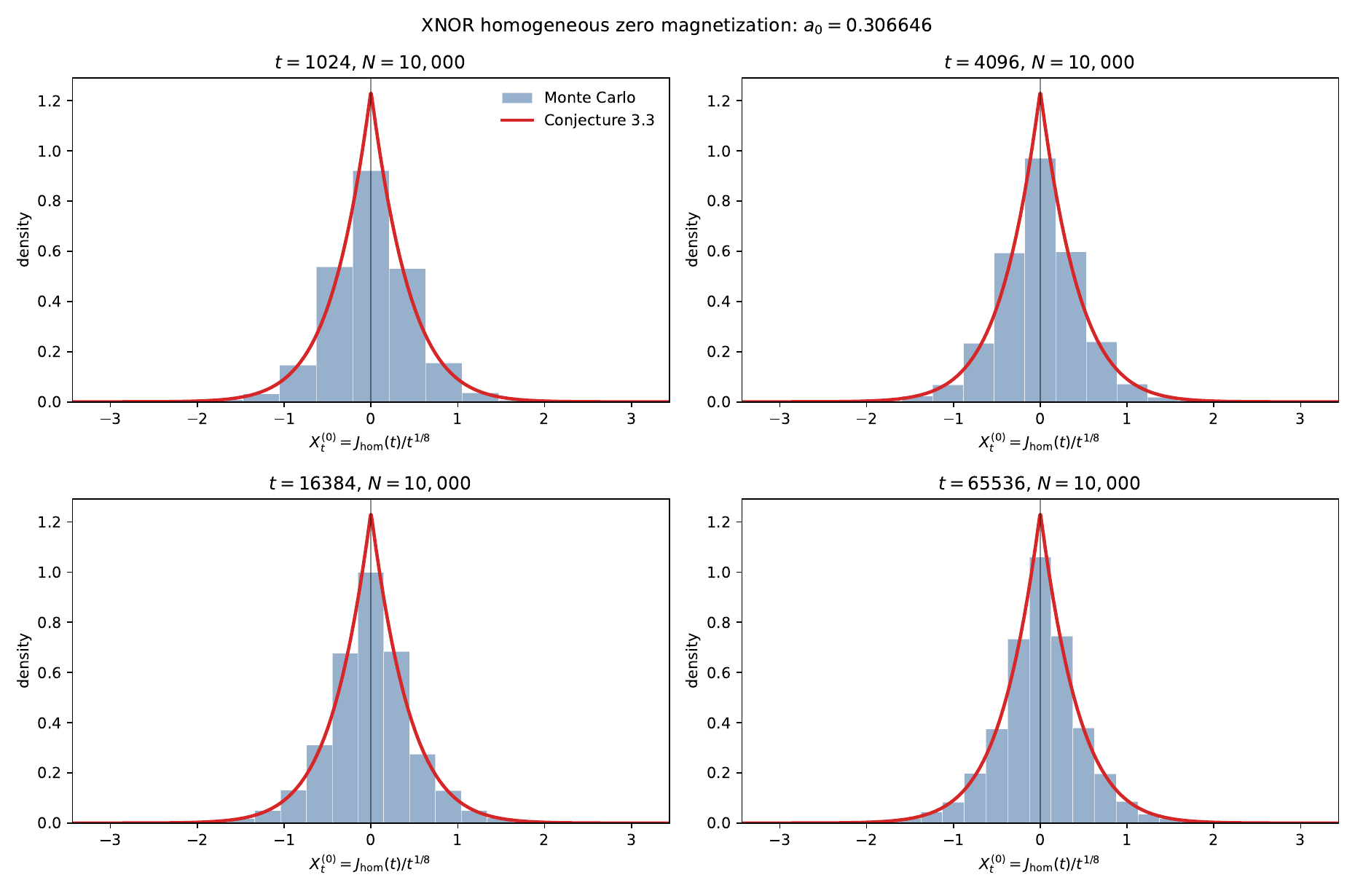}
\caption{Distribution of $\Jhom(t)/t^{1/8}$ in the main $L=512$ ensemble at
four selected times.  Histograms are the Monte Carlo data and the solid curve
is the M-Wright density $p_0$ from
\eqref{eq:main-results-m0-density}.}
\label{fig:numerics-zero-distributions}
\end{figure}

Both the distributional distance and the unscaled moment growth are consistent
with the proposed nested square-root scaling, with visible finite-time
corrections.
Figure~\ref{fig:numerics-zero-distributions} shows the direct density
comparison at four selected times.

We also tested the microscopic reduction leading to
\eqref{eq:phys-m0-brownian-limit}.  At the largest time the main ensemble gives
\begin{equation}
 \EE\!\left[\left(Q_t/t^{1/4}\right)^2\right]=0.36001,
 \qquad
 \EE\!\left[\left(N_t/t^{1/4}\right)^2\right]=0.09329,
 \qquad
 \frac{\EE[\Jhom(t)^2]}{\EE|N_t|}=1.01495.
 \label{eq:numerics-zero-diagnostics}
\end{equation}
The respective predictions are $2/(3\sqrt\pi)=0.37613$,
$1/(6\sqrt\pi)=0.09403$, and $1$.  The root-mean-square residual in
$N_t+Q_t/2$, divided by $t^{1/4}$, decreases monotonically from $0.2055$ to
$0.0907$ over the five times.  These observations independently support the
tagged Gaussian motion, the active-zero density reduction, and the zero-mean
colour sum that together produce the random-time Brownian limit.

\begin{figure}[H]
\centering
\includegraphics[width=\linewidth]{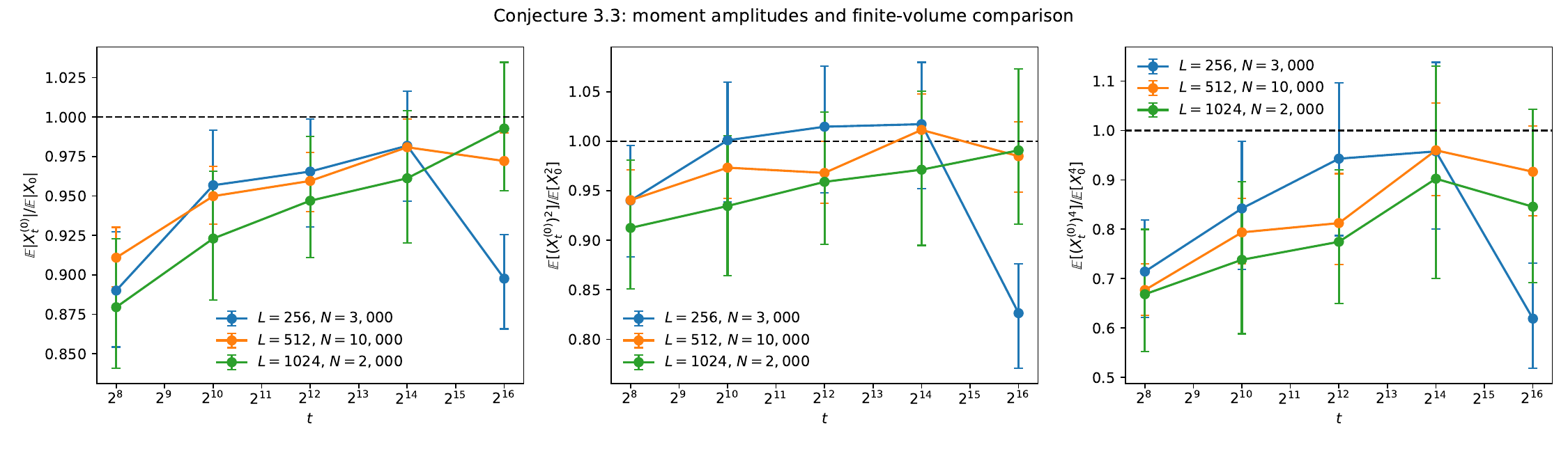}
\caption{Absolute first-, second-, and fourth-moment amplitudes divided by the
predictions of Conjecture~\ref{thm:main-summary-zero}.  The $L=512$ and
$L=1024$ results agree at $t=65536$ within Monte Carlo uncertainty; the
downturn for $L=256$ is consistent with a finite-volume effect.  Error bars are
95\% uncertainty intervals obtained by trajectory resampling.}
\label{fig:numerics-zero-volume-moments}
\end{figure}

A finite-volume check is particularly important at the largest time.  We ran
independent ensembles with $L=256$ and $L=1024$ through the same time grid.
Their final-time results are compared with the main run in
Table~\ref{tab:numerics-zero-volume} and
Figure~\ref{fig:numerics-zero-volume-moments}.

\begin{table}[!ht]
\centering
\small
\setlength{\tabcolsep}{4pt}
\begin{tabular}{r r cccc}
\hline
$L$ & samples & $\EE|X_t^{(0)}|$ & $\EE[(X_t^{(0)})^2]$ & $\EE[(X_t^{(0)})^4]$ & $W_1$ \\
\hline
$256$  & $3{,}000$  & $0.32608$ & $0.20219$ & $0.17464$ & $0.07258$ \\
$512$  & $10{,}000$ & $0.35315$ & $0.24096$ & $0.25851$ & $0.06268$ \\
$1024$ & $2{,}000$  & $0.36063$ & $0.24241$ & $0.23853$ & $0.06391$ \\
limit  &             & $0.36327$ & $0.24467$ & $0.28209$ & $0$ \\
\hline
\end{tabular}
\caption{Finite-volume comparison at $t=65536$.  The $L=512$ and $L=1024$
ensembles agree within sampling uncertainty, whereas the $L=256$ ensemble
shows a significant finite-volume deviation at this time.}
\label{tab:numerics-zero-volume}
\end{table}

For $L=1024$, the 95\% uncertainty intervals for all three moments overlap the
main-ensemble estimates and contain the predictions.  The difference between
the two fourth-moment point estimates is not statistically resolved by these
tail samples.  In contrast, the $L=256$ variance is inconsistent with both the
prediction and the central $L=512$ result.  The same smaller box agrees with
the larger volumes through $t=16384$, so its late decrease is consistent with
a boundary effect, rather than providing clear evidence against the conjecture.

Overall, the numerical results are consistent with
Conjecture~\ref{thm:main-summary-zero} over the accessible time and volume
ranges, although fourth-moment precision remains limited.

\section{Conclusions}

\label{sec:concl}

We formulated three explicit conjectures for the long-time current fluctuations in the XNOR hopping model. The conjectures
were supported by extensive numerical checks. A candidate proof is presented in the companion manuscript
\cite{PozsgayXNORProof2026}.

The XNOR hopping model is integrable, its stochastic Hamiltonian belongs to the family of Hamiltonians discussed in
\cite{PozsgayGomborHutsalyuk2021}. However, in this work we did not use the standard methods of integrability, such as
coordinate Bethe Ansatz. Instead, our main technique was specially adapted to this model: we used a sequence of
coordinate transformations that map the model to the well studied SSEP. If we are interested in computing other physical
properties of the XNOR hopping model, the mapping may not always be directly useful. In fact, a number of open problems
remain. For example, the exact time dependent magnetization profile that we get from the initial domain wall
distribution is not yet known. Asymptotic results could be computed using the methods of this work, but an exact formula
is not available. It would be desirable to continue work in this direction, and to derive further exact results for the model.

Another interesting future direction  is to extend these calculations to the folded XXZ quantum spin chain. The
kinetic term in that model has the same constraint as in the stochastic case. Yet, extending the computations to the
quantum mechanical model poses significant challenges. In this work the classical model was treated by estabilishing a direct
connection to tracer motion in SSEP. The same coordinate transformations that we used in this work map the folded XXZ
model to the Maassarani-Mathieu (MM) chain \cite{PozsgayEtAl2021}. In that model the motion of particles is dictated by the
free fermionic XX chain, whereas the dynamics of the ``colors'' of particles has the single-file and inert properties. 
The recent work \cite{FujimotoEtAl2026} already established  that the MM chain has anomalous current fluctuations.
Combining those results with the non-local mapping here is a promising direction. We leave this to future work.

\section*{Acknowledgments}

The author was supported by the Hungarian National Research,
Development and Innovation Office, NKFIH Grant No. K-145904.


\begin{thebibliography}{10}

\bibitem{KrajnikIlievskiProsen2022}
{\v{Z}}.~Krajnik, E.~Ilievski, and T.~Prosen, ``{Absence of Normal Fluctuations
  in an Integrable Magnet},'' {\em Phys. Rev. Lett.} {\bf 128} (2022)  090604,
  \href{http://arxiv.org/abs/2109.13088}{{\tt arXiv:2109.13088
  [cond-mat.stat-mech]}}.

\bibitem{KrajnikSchmidtPasquierIlievskiProsen2022}
{\v{Z}}.~Krajnik, J.~Schmidt, V.~Pasquier, E.~Ilievski, and T.~Prosen, ``{Exact
  Anomalous Current Fluctuations in a Deterministic Interacting Model},'' {\em
  Phys. Rev. Lett.} {\bf 128} (2022)  160601,
  \href{http://arxiv.org/abs/2201.05126}{{\tt arXiv:2201.05126
  [cond-mat.stat-mech]}}.

\bibitem{KrajnikSchmidtPasquierProsenIlievski2024}
{\v{Z}}.~Krajnik, J.~Schmidt, V.~Pasquier, T.~Prosen, and E.~Ilievski,
  ``{Universal anomalous fluctuations in charged single-file systems},'' {\em
  Phys. Rev. Research} {\bf 6} (2024)  013260,
  \href{http://arxiv.org/abs/2208.01463}{{\tt arXiv:2208.01463
  [cond-mat.stat-mech]}}.

\bibitem{KrajnikKlobasBertiniProsen2025}
{\v{Z}}.~Krajnik, K.~Klobas, B.~Bertini, and T.~Prosen, ``{Fluctuations of
  stochastic charged cellular automata},'' {\em J. Stat. Mech.} {\bf 2025}
  (2025)  053209, \href{http://arxiv.org/abs/2502.02509}{{\tt arXiv:2502.02509
  [cond-mat.stat-mech]}}.

\bibitem{McCullochDeNardisGopalakrishnanVasseur2023}
E.~McCulloch, J.~D. Nardis, S.~Gopalakrishnan, and R.~Vasseur, ``{Full Counting
  Statistics of Charge in Chaotic Many-Body Quantum Systems},'' {\em Phys. Rev.
  Lett.} {\bf 131} (2023)  210402, \href{http://arxiv.org/abs/2302.01355}{{\tt
  arXiv:2302.01355 [quant-ph]}}.

\bibitem{GopalakrishnanMcCullochVasseur2024}
S.~Gopalakrishnan, E.~McCulloch, and R.~Vasseur, ``{Non-Gaussian diffusive
  fluctuations in Dirac fluids},'' {\em Proc. Natl. Acad. Sci. U.S.A.} {\bf
  121} (2024)  e2403327121, \href{http://arxiv.org/abs/2401.05494}{{\tt
  arXiv:2401.05494 [cond-mat.stat-mech]}}.

\bibitem{McCullochVasseurGopalakrishnan2025}
E.~McCulloch, R.~Vasseur, and S.~Gopalakrishnan, ``{Ballistic modes as a source
  of anomalous charge noise},'' {\em Phys. Rev. E} {\bf 111} (2025)  015410,
  \href{http://arxiv.org/abs/2407.03412}{{\tt arXiv:2407.03412
  [cond-mat.stat-mech]}}.

\bibitem{Doyon2023}
B.~Doyon, G.~Perfetto, T.~Sasamoto, and T.~Yoshimura, ``{Ballistic macroscopic
  fluctuation theory},'' {\em SciPost Phys.} {\bf 15} (2023)  136,
  \href{http://arxiv.org/abs/2206.14167}{{\tt arXiv:2206.14167
  [cond-mat.stat-mech]}}.

\bibitem{YoshimuraKrajnik2025}
T.~Yoshimura and {\v{Z}}.~Krajnik, ``{Anomalous current fluctuations from Euler
  hydrodynamics},'' {\em Phys. Rev. E} {\bf 111} (2025)  024141,
  \href{http://arxiv.org/abs/2406.20091}{{\tt arXiv:2406.20091
  [cond-mat.stat-mech]}}.

\bibitem{FujimotoEtAl2026}
K.~Fujimoto, T.~Ishiyama, T.~Kurose, T.~Yoshimura, and T.~Sasamoto, ``{Exact
  Anomalous Current Fluctuations in Quantum Many-Body Dynamics},'' {\em arXiv
  e-prints} (2026)  , \href{http://arxiv.org/abs/2602.24008}{{\tt
  arXiv:2602.24008 [cond-mat.stat-mech]}}.

\bibitem{takato-stb-xxz-anomalous-fluct}
T.~{Yoshimura}, {\v{Z}}.~{Krajnik}, A.~{Bastianello}, and E.~{Ilievski},
  ``{Anomalous Hydrodynamic Fluctuations in the Quantum XXZ Spin Chain},''
  \href{http://dx.doi.org/10.1103/jrtg-yf1q}{{\em Phys. Rev. Lett.} {\bf 137}
  (2026) no.~8, 086302}, \href{http://arxiv.org/abs/2602.24242}{{\tt
  arXiv:2602.24242 [cond-mat.stat-mech]}}.

\bibitem{Zadnik2021}
L.~Zadnik and M.~Fagotti, ``{The Folded Spin-1/2 XXZ Model: I. Diagonalisation,
  Jamming, and Ground State Properties},'' {\em SciPost Phys. Core} {\bf 4}
  (2021)  010, \href{http://arxiv.org/abs/2009.04995}{{\tt arXiv:2009.04995
  [cond-mat.stat-mech]}}.

\bibitem{PozsgayEtAl2021}
B.~Pozsgay, T.~Gombor, A.~Hutsalyuk, Y.~Jiang, L.~Pristy{\'a}k, and E.~Vernier,
  ``{An integrable spin chain with Hilbert space fragmentation and solvable
  real time dynamics},'' {\em Phys. Rev. E} {\bf 104} (2021)  044106,
  \href{http://arxiv.org/abs/2105.02252}{{\tt arXiv:2105.02252
  [cond-mat.stat-mech]}}.

\bibitem{PozsgayGomborHutsalyuk2021}
B.~Pozsgay, T.~Gombor, and A.~Hutsalyuk, ``{Integrable hard rod deformation of
  the Heisenberg spin chains},'' {\em Phys. Rev. E} {\bf 104} (2021)  064124,
  \href{http://arxiv.org/abs/2108.13724}{{\tt arXiv:2108.13724
  [cond-mat.stat-mech]}}.

\bibitem{SalaEtAl2020}
P.~Sala, T.~Rakovszky, R.~Verresen, M.~Knap, and F.~Pollmann, ``{Ergodicity
  Breaking Arising from Hilbert Space Fragmentation in Dipole-Conserving
  Hamiltonians},'' {\em Phys. Rev. X} {\bf 10} (2020)  011047,
  \href{http://arxiv.org/abs/1904.04266}{{\tt arXiv:1904.04266
  [cond-mat.str-el]}}.

\bibitem{KhemaniHermeleNandkishore2020}
V.~Khemani, M.~Hermele, and R.~Nandkishore, ``{Localization from Hilbert space
  shattering: From theory to physical realizations},'' {\em Phys. Rev. B} {\bf
  101} (2020)  174204, \href{http://arxiv.org/abs/1910.01137}{{\tt
  arXiv:1910.01137 [cond-mat.stat-mech]}}.

\bibitem{MoudgalyaMotrunich2022}
S.~Moudgalya and O.~I. Motrunich, ``{Hilbert Space Fragmentation and Commutant
  Algebras},'' {\em Phys. Rev. X} {\bf 12} (2022)  011050,
  \href{http://arxiv.org/abs/2108.10324}{{\tt arXiv:2108.10324
  [cond-mat.stat-mech]}}.

\bibitem{MenonBarmaDhar1997}
G.~I. Menon, M.~Barma, and D.~Dhar, ``{Conservation laws and integrability of a
  one-dimensional model of diffusing dimers},'' {\em Journal of Statistical
  Physics} {\bf 86} (1997)  1237--1263,
  \href{http://arxiv.org/abs/cond-mat/9703059}{{\tt arXiv:cond-mat/9703059
  [cond-mat.stat-mech]}}.

\bibitem{SinghWareVasseurFriedman2021}
H.~Singh, B.~A. Ware, R.~Vasseur, and A.~J. Friedman, ``{Subdiffusion and
  Many-Body Quantum Chaos with Kinetic Constraints},'' {\em Phys. Rev. Lett.}
  {\bf 127} (2021)  230602, \href{http://arxiv.org/abs/2108.02205}{{\tt
  arXiv:2108.02205 [cond-mat.stat-mech]}}.

\bibitem{FeldmeierWitczakKrempaKnap2022}
J.~Feldmeier, W.~Witczak-Krempa, and M.~Knap, ``{Emergent tracer dynamics in
  constrained quantum systems},'' {\em Phys. Rev. B} {\bf 106} (2022)  094303,
  \href{http://arxiv.org/abs/2205.07901}{{\tt arXiv:2205.07901
  [cond-mat.str-el]}}.

\bibitem{GopalakrishnanMorningstarVasseurKhemani2024}
S.~Gopalakrishnan, A.~Morningstar, R.~Vasseur, and V.~Khemani, ``{Distinct
  universality classes of diffusive transport from full counting statistics},''
  {\em Phys. Rev. B} {\bf 109} (2024)  024417,
  \href{http://arxiv.org/abs/2203.09526}{{\tt arXiv:2203.09526
  [cond-mat.stat-mech]}}.

\bibitem{GopalakrishnanVasseur2023}
S.~Gopalakrishnan and R.~Vasseur, ``{Anomalous transport from hot
  quasiparticles in interacting spin chains},''
  \href{http://dx.doi.org/10.1088/1361-6633/acb36e}{{\em Rep. Prog. Phys.} {\bf
  86} (2023)  036502}, \href{http://arxiv.org/abs/2208.11133}{{\tt
  arXiv:2208.11133 [cond-mat.str-el]}}.

\bibitem{PozsgayXNORProof2026}
B.~Pozsgay, ``{Current fluctuations in the stochastic XNOR hopping process (An
  AI-generated candidate proof)}.'' Zenodo, July, 2026.
\newblock \url{https://doi.org/10.5281/zenodo.21496322}.

\bibitem{Leiden}
J.~Alper, M.~J. Barany, A.~C. Villarello, S.~Dahmen, W.~Dean, K.~Ganapathy,
  M.~Harris, D.~Holmes, M.~Jamnik, S.~Kelk, B.~Kra, U.~Martin,
  B.~Naskr{\k{e}}cki, R.~Ochigame, J.~Portegies, and J.~Schmitt, ``{Leiden
  Declaration on Artificial Intelligence and Mathematics}.'' Zenodo, June,
  2026.
\newblock \url{https://doi.org/10.5281/zenodo.20302944}.

\bibitem{Spohn1991}
H.~Spohn, {\em {Large Scale Dynamics of Interacting Particles}}.
\newblock Springer, 1991.

\bibitem{KipnisLandim1999}
C.~Kipnis and C.~Landim, {\em {Scaling Limits of Interacting Particle
  Systems}}.
\newblock Springer, 1999.

\bibitem{DerridaGerschenfeld2009}
B.~Derrida and A.~Gerschenfeld, ``{Current Fluctuations of the One Dimensional
  Symmetric Simple Exclusion Process with Step Initial Condition},''
  \href{http://dx.doi.org/10.1007/s10955-009-9772-7}{{\em Journal of
  Statistical Physics} {\bf 136} (2009) no.~1, 1--15},
  \href{http://arxiv.org/abs/0902.2364}{{\tt arXiv:0902.2364
  [cond-mat.stat-mech]}}.

\bibitem{Arratia1983}
R.~Arratia, ``{The motion of a tagged particle in the simple symmetric
  exclusion system on {$\mathbb Z$}},'' {\em Annals of Probability} {\bf 11}
  (1983)  362--373.

\bibitem{PeligradSethuraman2008}
M.~Peligrad and S.~Sethuraman, ``{On fractional Brownian motion limits in
  one-dimensional nearest-neighbor symmetric simple exclusion},'' {\em ALEA
  Latin American Journal of Probability and Mathematical Statistics} {\bf 4}
  (2008)  245--255, \href{http://arxiv.org/abs/0711.0017}{{\tt arXiv:0711.0017
  [math.PR]}}.

\bibitem{Gillespie1976}
D.~T. Gillespie, ``{A General Method for Numerically Simulating the Stochastic
  Time Evolution of Coupled Chemical Reactions},''
  \href{http://dx.doi.org/10.1016/0021-9991(76)90041-3}{{\em Journal of
  Computational Physics} {\bf 22} (1976) no.~4, 403--434}.

\end{thebibliography}

\providecommand{\href}[2]{#2}\begingroup\raggedright\endgroup

\end{document}